\documentclass[twocolumn,trackchanges]{aastex701}

\newcommand{\alf}{Alfv\'enic }
\newcommand{\nalf}{Alfv\'enic}
\newcommand{\ma}{${\cal{M}}_A$}
\newcommand{\sma}{${\cal{M}}_A$ }

\received{TBD}
\revised{TBD}
\accepted{TBD}
\submitjournal{The Astrophysical Journal Supplement Series}
\shorttitle{Disentangling Magnetic Field Geometry}
\shortauthors{Kressy et al.}

\usepackage{graphicx} 
\usepackage{amsmath}
\usepackage{natbib}
\usepackage{microtype}

\begin{document}

\title{Disentangling 3D Magnetic Field Geometry from Turbulence: A Polarization-Based Classification Method}

\author[orcid=0000-0002-3469-5774,sname='Kressy']{Sophia S. Kressy}
\affiliation{Department of Physics and Astronomy, University of North Carolina Chapel Hill, Chapel Hill, NC 27599}
\email[show]{skressy@email.unc.edu}  

\author[orcid=0000-0000-0000-0000,sname='Heitsch']{Fabian Heitsch}
\affiliation{Department of Physics and Astronomy, University of North Carolina Chapel Hill, Chapel Hill, NC 27599}
\email[show]{fheitsch@email.unc.edu}  

\keywords{\uat{Interstellar magnetic fields}{845} --- \uat{Polarimetry}{1278} --- \uat{Star formation}{1569} --- \uat{Computational methods}{1965}}

\begin{abstract}
We develop a decision tree classifier that recovers 3D magnetic field geometry from synthetic polarization and position angle maps of a magnetized, turbulent field.
We test the classifier for a series of geometries: Uniform, Wavy, Helical, and Hourglass.
The classifier is calibrated across a range of \alf Mach numbers and injected Stokes Q,U noise levels for each geometry.
Our models show the median and skewness of the polarization fraction, as well as the position angle circular variance, vary systematically with both geometry and the \alf Mach number.
We find that the 3D geometry can be isolated by using a variety of polarization and position angle metrics to break degeneracies.
The classifier recovers most geometries well for sub-\alf cases (\ma$<1$).
For \ma $>1$ polarization and position angle statistics converge across all geometries and the classifier fails to recover the magnetic field structure.
\end{abstract}

\section{Introduction}
\label{sec:intro}
The interstellar medium (ISM) is both turbulent and magnetized \citep{Ferriere+2001, Elmegreen2004}.
Understanding the role turbulence and magnetic fields play in star formation has been an ongoing effort \citep{Maclow2004,McKee2007,Ballesteros+2007}. 
Simple gas-consumption arguments overestimate the Galactic star formation rate by orders of magnitude \citep{Scalo+1986, KennicuttEvans+2012} and magnetic fields have historically been considered a leading candidate for this inefficiency \citep{Shu+1987, Adams87, Lizano87}.
Under ideal magnetohydrodynamic (MHD) conditions the magnetic field is ``frozen'' to the gas, and, if sufficiently strong, can provide support against gravitational collapse.
This is quantified by the mass-to-flux ratio \citep{Shu+1987}: subcritical clouds can be magnetically supported against collapse, while supercritical clouds will collapse.

There has been substantial debate about the extent to which magnetic fields control star formation \citep{Ballesteros+2007, Crutcher+2009}.
Zeeman measurements across various column densities, and therefore environments, have shown us that magnetic fields remain an important contributor in the transition from diffuse to dense gas \citep{Crutcher+1999, Crutcher+2012}. 
How exactly this transition occurs remains largely unanswered, though scenarios have been proposed \citep{Wang+2020} and simulations are beginning to cover the relevant spatial and dynamical range \citep{Seifried+2020}. 
In a scenario where clouds form due to the sweep-up of material over larger scales, the development of star-forming, i.e. supercritical, clouds may just be the result of a selection effect \citep{Hartmannetal2001,Heitsch+2009,InoueInutsuka+2009}.

Beyond field strength, magnetic field \emph{geometry} informs us of where gas may flow, since the gas and field are dynamically coupled in a turbulent, magnetized plasma.
Studies of filaments, elongated structures of gas and dust, have shown that their orientation is related to the local magnetic field \citep[e.g.,][]{Hartmann2002, Palmeirim+2013, Seifried+2020}.
\citet{Planck+2015}'s all-sky survey revealed a strong tendency for filaments to be oriented either parallel or perpendicular to the local POS field depending on column density (see also \citealt{Soler+2017a} for the Vela C molecular complex). 
This relationship has been seen in MHD simulations as well \citep{Soler+2017b, Hennebelle+2013}.

Molecular clouds are considered to harbor supersonic turbulence on scales $\geq 0.1$~pc.
Turbulence plays a central role in seeding density fluctuations from which cores form \citep[see reviews by][]{Maclow2004, Ballesteros2007, McKee2007}. 
On the other hand, it has also been proposed as a mechanism to support clouds against gravitational collapse \citep{Bonazzola1987, Bonazzola1992, Krumholz2005, Hennebelle2008}.
However, this picture is complicated by the rapid dissipation of supersonic turbulence, which must be continuously replenished to sustain support.

The strength of turbulence can be quantified by the \alf Mach number, $\mathcal{M}_A \equiv v / v_A$, where $v_A = B/ \sqrt{4\pi\rho}$ is the Alfvén speed. 
The \alf Mach number is a proxy for the ratio of turbulent to magnetic energy. 
For $\mathcal{M}_A < 1$ (sub-\nalf) the magnetic field is dynamically dominant, while for $\mathcal{M}_A > 1$ (super-\nalf) turbulence dominates and the magnetic field is dragged with it.
$\mathcal{M}_A$ is not directly observable, but is inferred from velocity and magnetic field tracers.

ISM turbulence is magnetized, but also anisotropic and compressible \citep{Armstrong+1995, Elmegreen2004, McKee2007}.
\citet{Goldreich+1995} (hereafter GS95) developed a magnetized Kolmogorov spectrum of velocity and a scale-dependent anisotropy for incompressible MHD, which has been supported by numerical simulations \citep{Cho+2000, Maron+2001}.
\citet{Cho+2003} (hereafter CL03) extend the GS95 model to add compressibility \citep[e.g.,][]{Falceta+2008}.
GS95 and CL03 provide a theoretical base to model the ISM and observational tracers, which we will adopt for the purpose of this study.

Magnetic field strengths in cold gas can be inferred via the Zeeman effect in OH and HI \citep[e.g.,][]{Crutcher+2012}, and via the Davis-Chandrasekhar-Fermi method using polarimetry and line-of-sight velocity dispersions \citep{ChandrasekharFermi+1953}. 
Magnetic field orientation in the plane-of-sky can be traced by polarimetry in emission or absorption \citep{Hall+1949, Hiltner+1949, Davis+1951, Hildebrand+1988}.

Extensive polarimetry datasets have been assembled from ground-based and space-based facilities alike, including \textit{Planck} \citep{Planck+2015}.
Higher-resolution datasets from instruments such as NASA's Stratospheric Observatory for Infrared Astronomy (SOFIA) \citep{SOFIA2018, Chuss2019} and the B-fields In STar-forming Region Observations (BISTRO) survey \citep{Ward+2017} have further resolved magnetic field structure within individual star-forming regions.

Despite the various methods for observation, inferring the 3D field geometry has proven difficult. 
Polarimetry has no way of accessing the LOS component, and methods that do measure LOS field components operate at different scales and make a variety of assumptions (e.g., Faraday assumes electron densities). 
Various field geometries have been invoked to explain filament stability and polarimetry observations \citep{FiegePudritz2000a, FiegePudritz2000b, FiegePudritz2000c, Myers+2020}, but a unified, quantitative picture connecting geometry to observable polarization statistics is needed.

Forward-modeling approaches have been successful in developing realistic synthetic polarization maps from MHD simulations, from a variety of initial conditions \citep{FiegePudritz2000a, Falceta+2008, Reissl+2018, Reissl+2021}, against which observations can be compared. 

Other studies have used indirect approaches via polarization, Zeeman or DCF statistics derived from MHD simulations \citep{Burkhart+2012, Chen+2019, Shane+2024, Hoang+2024} to infer magnetic field geometry or the inclination (LOS) angle of the magnetic field. 
This approach allows for direct measurements from observations, rather than a comparative analysis.

In this study, we use a combination of the two approaches. 
We forward-model a series of magnetic field geometric templates for a variety of turbulence strengths.
We then create synthetic polarization data, which we analyze for trends indicative of the underlying field geometries. 

Diagnosing magnetic field geometry in molecular clouds has primarily relied on POS polarimetry, as dust emission and absorption are less time-intensive and more extensive than Zeeman measurements.
While prior work has linked polarization statistics to the \alf Mach number in general turbulence simulations 
\citep[e.g.,][]{Burkhart+2012, Hu+2023}, it remains largely untested whether statistical properties of polarization observables can serve as a geometry diagnostic. 
However, the observed polarization degree (P) mixes contributions from turbulence, large-scale field geometry, grain-alignment efficiency, and depolarization from LOS integration, making it difficult to isolate the LOS magnetic field's contribution to the signal.
Disentangling these contributions is not straightforward and requires multiple observables: Zeeman strength, velocity profile, and polarimetry. 

This paper investigates whether the magnetic field geometry can be revealed using polarimetry information alone.
The models implement the anisotropic incompressible turbulence of \citet{Goldreich+1995} and \citet{Cho+2003} for four magnetic field geometries: Uniform, Wavy, Helical, and Hourglass.
The templates were chosen because they have been modeled and/or observed in prior work, and they represent a physically motivated sequence from ordered to complex morphology.

We find that the skewness of the polarization fraction distribution varies systematically with $\mathcal{M}_A$, and that this relationship is geometry-dependent.
Some geometries can be distinguished by their polarization fraction distributions alone; remaining degeneracies can be broken with a variety of polarization statistics: polarization fraction median ($\rm \bar P$) and peak ($\rm P_{peak}$), circular variance in PA ($\rm CV_{PA}$), number of peaks in PA distribution ($N_{PA_{peaks}}$), and the mass-to-flux ($M/\Phi$) ratio. 
A decision tree logic can be employed to break all degeneracies and recover the input geometry.
We find that in the absence of grain alignment effects, the decision tree recovers the underlying geometry for uncertainty levels in Stokes U and Q ranging from 0-10\%. A future contribution will address grain alignment effects. 

Section~\ref{sec:methods} describes the model construction, magnetic field geometries and turbulence prescription.
Section~\ref{sec:results} explores the polarization statistics used to recover the field. 
These statistics then inform Section~\ref{sec:decisiontree}, which discusses the decision tree and classification method, as well as degeneracies.
Section~\ref{sec:diss} discusses and contextualizes results.
Main conclusions are summarized in Section~\ref{sec:sum}.

\section{Model Framework}
\label{sec:methods}
Polarization maps are generated assuming one of the four underlying 3D field geometries. 
We add the effects of turbulence by perturbing the field vectors following \citet{Goldreich+1995} (GS95), \citet{Cho+2003} (CL03) and \citet{Lazarian+2006}, preserving the $\nabla \cdot B=0$ constraint. 
Stokes parameters are then derived by integrating the 3D field components along the LOS. 
We assume constant emissivity, an optically thin gas, and 100\% grain alignment.

\subsection{Magnetic Field Geometries}
A series of magnetic field templates common to theory and observations -- Uniform, Wavy, Helical, and Hourglass -- was developed.
Figure~\ref{fig:pol_maps} shows all geometric templates in the XY-plane for \ma$=0.01$ and $\sigma_{Q,U}=0$ to favor geometry visualization.
The governing equations of the geometries are described as below.

\subsubsection{Uniform}
Uniform fields, often associated with diffuse regions where the magnetic field is dynamically dominant, have been observed in dark and translucent clouds \citep{Goodman+1992, Panopoulou+2016}.
Figure~\ref{fig:pol_maps} shows the synthetic PA map overlaid on the polarization fraction map for a Uniform field.
Integration is along the Z-axis, with the POS as the XY-plane. 
The orientation angle of the field is an arbitrary, user-defined variable. 
In Figure~\ref{fig:pol_maps} the orientation angle is set to 45 degrees and the B$_z$ component is set to 0.
We assume B$_z = 0$ for our models. 

\subsubsection{Wavy}
Wavy fields represent a perturbation of a Uniform field, as would arise from an Alfv\'{e}n wave propagating along or transverse to the mean field, or from a large-scale shock \citep{Boekholt+2017, Schleicher+2018, Stutz+2016}.
A Wavy magnetic field template is generated following the equations,
\begin{align}
    B_x &= b_x \\
    B_y &= A\cos{xk} \\
    B_z &= b_z .
\end{align}
B$_y$ contains the oscillation of the cosine wave, $A$ is the amplitude, and k is the wavenumber. The B$_x$ and B$_z$ components are defined by the user. 
Figure~\ref{fig:pol_maps} shows the synthetic polarization map for a Wavy field for B$_x=1$, B$_z=0$, $A=1$, and wavenumber $k=1$, evaluated over the spatial domain $x\in [0,2\pi]$, which we adopt for our modeling.

\subsubsection{Helical}
Helical fields have been studied analytically and numerically \citep{FiegePudritz2000a, FiegePudritz2000b, Matthews+2001, Tahani+2018} and have been invoked to explain observed depolarization \citep{Matthews+2001}.
A Helical field can be described by the following equations,
\begin{align}
    r &= \sqrt{x^2 + y^2} \\
    B_x &= (-Y\cos{\alpha})/r \\
    B_y &=(X\cos{\alpha})/r \\
    B_z &= \sin{\alpha} ,
\end{align}
where $\alpha$ is the pitch angle of the helix. 
Figure~\ref{fig:pol_maps}, Helical Axis = 1 shows the Helical structure from the side, in the XY plane, for $\alpha = 0$.
The Helical template can also be viewed, or integrated, down the `barrel' (Y-axis), which is seen in Helical Axis = 2. 

\subsubsection{Hourglass}
Hourglass fields may arise from gravitational ``pinching'' of an initially ordered field.
\citet{MestelSpitzer1956} and \citet{Mouschovias1976a, Mouschovias1976b} first suggested the geometry in the context of gravitational collapse of magnetic interstellar clouds.
\citet{GalliShu1993a, GalliShu1993b} then approached it with a semi-analytical and numerical treatment.
As material collapses toward a central concentration, the field is dragged with the material, producing the characteristic Hourglass pinch.
This morphology has been detected in several star-forming cores, including OMC-1 \citep{Pattle+2017, Ward+2017} and around the low-mass protostar L1157-mm \citep{Stephens+2013}.
It has also been modeled in the protostellar context by \citet{Basu+2024}.

Here, we use the normalized, axis-symmetrical poloidal version described by \citet{Basu+2024}, which builds on \citet{Ewertowski+2013}, in cylindrical coordinates,
\begin{align}
    \tilde{B}_r &= \sum_{m=1}^{\infty} \beta_mJ_1(a_{m,1}\tilde{r})[\rm erfc(\frac{a_{m,1}\eta}{2} - \frac{\tilde{z}}{\eta})\textit{e}^{-a_{m,1}\tilde{z}} \\ 
    &- erfc(\frac{a_{m,1}\eta}{2} + \frac{\tilde{z}}{\eta})\textit{e}^{a_{m,1}\tilde{z}}] \\
    \tilde{B}_z &= \sum_{m=1}^{\infty} \beta_m J_0(a_{m,1}\tilde{r})[\rm erfc(\frac{a_{m,1}\eta}{2} + \frac{\tilde{z}}{\eta})\textit{e}^{a_{m,1}\tilde{z}} \\
    &+ erfc(\frac{a_{m,1}\eta}{2} - \frac{\tilde{z}}{\eta})\textit{e}^{-a_{m,1}\tilde{z}}] + 1.
\end{align}
$\tilde{B_r} = B_r/B_0$ and $\tilde{B_z} = B_z/B_0$ are the normalized radial and azimuthal components to the magnetic field, such that $B_{poloidal} = B_r\hat{r} + B_z\hat{z}$. 
$J_0(x)$ is the Bessel function of the 0th order. $\lambda_m$ is defined by the eigenvalues,
\begin{align}
    \lambda_m = (\frac{a_{m,1}}{R})^2
\end{align}
where $a_{m,1}$ is the $m^{th}$ positive root of $J_1(x)$, the Bessel function of the 1st order. $R$ is the outer radius of the solution. 
$\eta$ is the coefficient defined by $\eta = h/R$, where $h$ is a free parameter. $\tilde{r}$ and $\tilde{z}$ are the normalized components by $R$.
$B_0$ is the magnetic field strength in the Z direction. $\beta_m$ is defined as,
\begin{align}
    \beta_m = B_m/B_0 = k_m\sqrt{\lambda_m}/B_0
\end{align}
where $B_m$ is in the units of the magnetic field. $k_m$ are some unknown fitting coefficients. In this computational case, $\beta_m$ is assumed to be a user-defined coefficient ($\approx$1).
The first three terms of the function series were taken when finding $\tilde{B_r}$ and $\tilde{B_z}$, as per suggestion by \citet{Basu+2024}.
Although \citet{Basu+2024} provide a $B_{\phi}$ component to introduce twisting of the Hourglass field, in this study it was forgone to favor simplicity.
Figure~\ref{fig:pol_maps}, Hourglass Axis = 1 shows the Hourglass template facing the `barrel' in the XY-plane. 
Similar to the Helical field template, the viewing or integration axis of the Hourglass template can be modified to look `down the barrel' and consider different vantage points.
Figure~\ref{fig:pol_maps}, Hourglass Axis = 2, shows down the Hourglass `barrel', along Y-axis integration.
Hourglass Axis = 2 is almost entirely in the LOS, and therefore shows all depolarization in the POS.

\subsection{Turbulence}
Instead of using MHD turbulence simulations to generate polarization maps \citep[e.g.,][]{Ostriker+2001, Heitsch+2001}, we implement turbulence as a prescription for perturbing the polarization angles around the local mean field direction. This choice provides us with full control of the test models, and allows us to test magnetic field geometries that otherwise would be challenging to reproduce in full simulations.

To model the whole range of \alf Mach numbers, the turbulence prescription needs to cover the sub- and super-\alf environments with anisotropy.
For anisotropic turbulence, we adopt the scaling relations outlined by CL03, constructing a scale-dependent spectral blend.

For sub-\alf cases (see Section~\ref{sec:subA}), CL03 found that turbulence is weak at large scales and transitions to strong GS95 turbulence at small scales, with the transition governed by the perpendicular wavenumber $k_{\perp}$ . 

For super-\alf cases (see Section~\ref{sec:superA}), turbulence is hydrodynamic at large scales and transitions to GS95 at small scales, with the transition governed by the isotropic wavenumber $k$ .

The turbulent field is constructed in Fourier space. In each realization, complex Gaussian random amplitudes are drawn, producing a random phase field in Fourier space. 
The scale-dependent power spectrum is then imposed as a spectral weight. 
The field is made divergence-free via a solenoidal projection, also performed in Fourier space. 
Finally, the resulting field is transformed back to real space via an inverse Fourier transform.

\subsubsection{Sub-\alf} \label{sec:subA}
For the sub-\alf (\ma$< 1$) we follow the treatment of GS95, CL03, and \citet{Lazarian+2006}. 
In this environment, turbulence is anisotropic from the injection scale $L$. 
At large scales, the turbulence is weak: wave packets preserve their parallel coherence length ($l_{\parallel}$) while developing structure perpendicular to the magnetic field, causing the perpendicular coherence length ($l_{\perp}$) to decrease. 
This weak cascade does not persist indefinitely; critical balance is reached at the transition scale $l_{trans}$, defined by the condition that the wave crossing time equals the nonlinear interaction time: $l_{trans} \approx LM_A^2$.
Here $l_{trans}$ marks the weak-to-strong turbulence transition.
Above this scale the cascade is weak, below it the turbulence enters the strong GS95 regime. 
Below $l_{trans}$, the turbulence follows GS95 scalings.
For ${\cal{M}}_A < 1$, the magnetic field fluctuates about a well-defined mean direction. 
The transition wavenumber corresponding to $l_{trans}$ is:
\begin{align}
    k_{trans} &\approx \frac{2\pi}{L} M_A^{-2}
\end{align}

\subsubsection{Super-\alf} \label{sec:superA}
For super-\alf, ${\cal{M}}_A > 1$, turbulence follows the isotropic hydrodynamic Kolmogorov cascade from the injection scale $L$ down to the Alfv\'en scale $l_A$: $l_A \approx LM_A^{-3}$
where $l_A$ is the scale at which the magnetic field becomes dynamically important, i.e., when $\nu_l = V_A$. Unlike $l_{trans}$ in the sub-Alfv\'enic case, which marks a weak-to-strong transition within a magnetized cascade, $l_A$ marks the scale at which magnetic forces first become relevant, setting the injection scale for GS95 turbulence. 
Below $l_A$, eddies become elongated along the local magnetic field 
direction, with $\nu_l \approx V_A(l_{\perp}/l_A)^{1/3}$. 
The eddy anisotropy is characterized by: $l_{\parallel} \approx L\left(\frac{l_{\perp}}{L}\right)^{2/3} M_A^{-1}$. 
Thus, while turbulence remains isotropic at scales $l > l_A$, it develops 
$(l_{\perp}/l_A)^{1/3}$ anisotropy for $l < l_A$. 
The corresponding transition wavenumber is:
\begin{align}
    k_A &\approx \frac{2\pi}{L}M_A^3
\end{align}

\subsubsection{Spectral Weights}
For each turbulent environment, we apply a spectral weight $W(k)$ derived from the corresponding energy spectrum. 
In 3D Fourier space, complex amplitudes with magnitude $W(k)$ define the power spectrum as:
\begin{equation}
    P_{3D} \approx W(k)^2
\end{equation}
Relating the 1D energy spectrum $E(k)$ to $P_{3D}(k)$ by integrating over a spherical shell of radius $k$:
\begin{align}
    E(k) &= P_{3D}(k) \cdot 4\pi k^2 \implies P_{3D}(k) \approx \frac{E(k)}{4\pi k^2} \\
    W(k) &\approx E(k)^{1/2} \, k^{-1}
\end{align}
This derivation assumes spherical shell geometry and an isotropic field; no analogous integration scheme for anisotropic 3D energy spectra is currently adopted.

\paragraph{$W_{\rm hydro}$}
For the hydrodynamic regime, we assume an isotropic Kolmogorov energy spectrum, $E(k) = k^{-5/3}$, which gives a spectral weight of,
\begin{align}
    W_{\rm hydro}(k) = k^{-11/6}.
\end{align}

\paragraph{$W_{\rm weak}$}
In the weak turbulence regime, energy transfer is predominantly perpendicular, with $k_{\perp} \gg k_{\parallel}$. 
Following \citet{Lazarian+1999}: Appendix A, Eq.~2, the nonlinear energy transfer rate scales as $\nu_k \propto k_{\perp}^{-1/2}$. 
Using $E(k) \approx \nu_k^2/k$ from \citet{Monin+1975}, this yields, 
\begin{align}
    W_{\rm weak}(k_{\perp}) = k_{\perp}^{-2}.
\end{align}
We note that while GS95 refers to this as an intermediate turbulence regime, the underlying physics is consistent with CL03's weak turbulence treatment.

\paragraph{$W_{\rm GS95}$}
For GS95, the Alfv\'en mode energy spectrum scales as $E_A(k) \propto k_{\perp}^{-5/3}$ \citep{Cho+2003}, giving $W(k_{\perp}) = k_{\perp}^{-11/6}$. 
To enforce the scale-dependent anisotropy of GS95, in which $k_{\parallel} \propto k_{\perp}^{2/3}$, we apply a Gaussian filter in the parallel direction:
\begin{equation}
    W_{\rm GS95}(k) = k_{\perp}^{-11/6} \exp\!\left[-\left(\frac{k_{\parallel}}
    {k_{\perp}^{2/3}}\right)^2\right]
\end{equation}
This weight encompasses both Alfv\'en and slow modes. Fast modes are excluded because, without solving the MHD equations, specifying the mode energy fractions would require additional assumptions.
Additionally, fast modes contribute primarily isotropic small-scale fluctuations that would suppress the overall polarization fraction without significantly affecting the experimental outcomes.

\subsection{Synthetic Stokes Maps}
For a given magnetic field model, we follow \citet{Zweibel+1996}. 
The line-of-sight is taken along the z-axis. 
For a single line-of-sight, the Stokes parameters are given by, 
\begin{equation} \label{eq:pol_long}
\begin{split}
    P &= Q+iU \\
    &= \frac{1}{\int n(z)dz} \int f(z) n(z) \frac{(B_x+iB_y)^2}{B_x^2+B_y^2} \cos^2\gamma \,dz,
\end{split}
\end{equation}
where $n(z)$ is the local density and $f(z)$ describes additional physical effects such as grain alignment.
We assume a constant density and physical effects to be negligible or constant.

For such assumptions, Stokes U, Q, and $\cos^2\gamma$ are defined by the magnetic field components,
\begin{align}
    \cos^2\gamma &= \frac{B_x^2+B_y^2}{B_x^2+B_y^2+B_z^2} \label{eq:cos}\\
    U &= \int\frac{2B_xB_y}{B_x^2+B_y^2}\cos^2\gamma \,dz \label{eq:U}\\
    Q &= \int \frac{B_y^2-B_x^2}{B_x^2+B_y^2}\cos^2\gamma \,dz\label{eq:Q}.
\end{align}
U and Q are then integrated along the LOS and can change axis (X,Y,Z) depending on the geometry -- i.e. Hourglass and Helical both produce very different polarization maps depending on the integration axis. If the integration axis is changed, so will the B components in the above expressions. 

Noise ($\sigma_{Q,U}$) can be injected into the Stokes parameters through intensity $I$,
\begin{align}
    I_{obs} &= \int_z dl \\
    \sigma_I &= \sigma_{Q,U} I_{obs}.
\end{align}
So that Stokes U and Q become,
\begin{align}
    Q_{obs} &= Q_{int} + N(0,\sigma_I) \\
    U_{obs} &= U_{int} + N(0,\sigma_I)
\end{align}
where $N(0,\sigma_I)$ is drawn from a normal distribution with mean equal to 0 and a standard deviation of $\sigma_I$ -- the scaled injected noise.
The polarization fraction is found with,
\begin{align}
    P = \frac{\sqrt{Q_{obs}^2+U_{obs}^2}}{I_{obs}}
\end{align}
However, because the P$_{obs}$ is a positive-definite quantity, noise biases it upward, so Rice Debiasing \citep{Serkowski+1958, Wardle+1974} is implemented for $\sigma_{P}>0$,
\begin{align}
    P_{debiased} = \sqrt{P^2 - \sigma_{Q,U}^2}
\end{align}
Finally, the position angle (PA) is defined as,
\begin{align}
    \phi = \frac{1}{2}\tan^{-1}\frac{U_{obs}}{Q_{obs}} . \label{eq:phi_uq}
\end{align}

\begin{figure}[t]
\centering
\includegraphics[width=0.38\paperwidth]{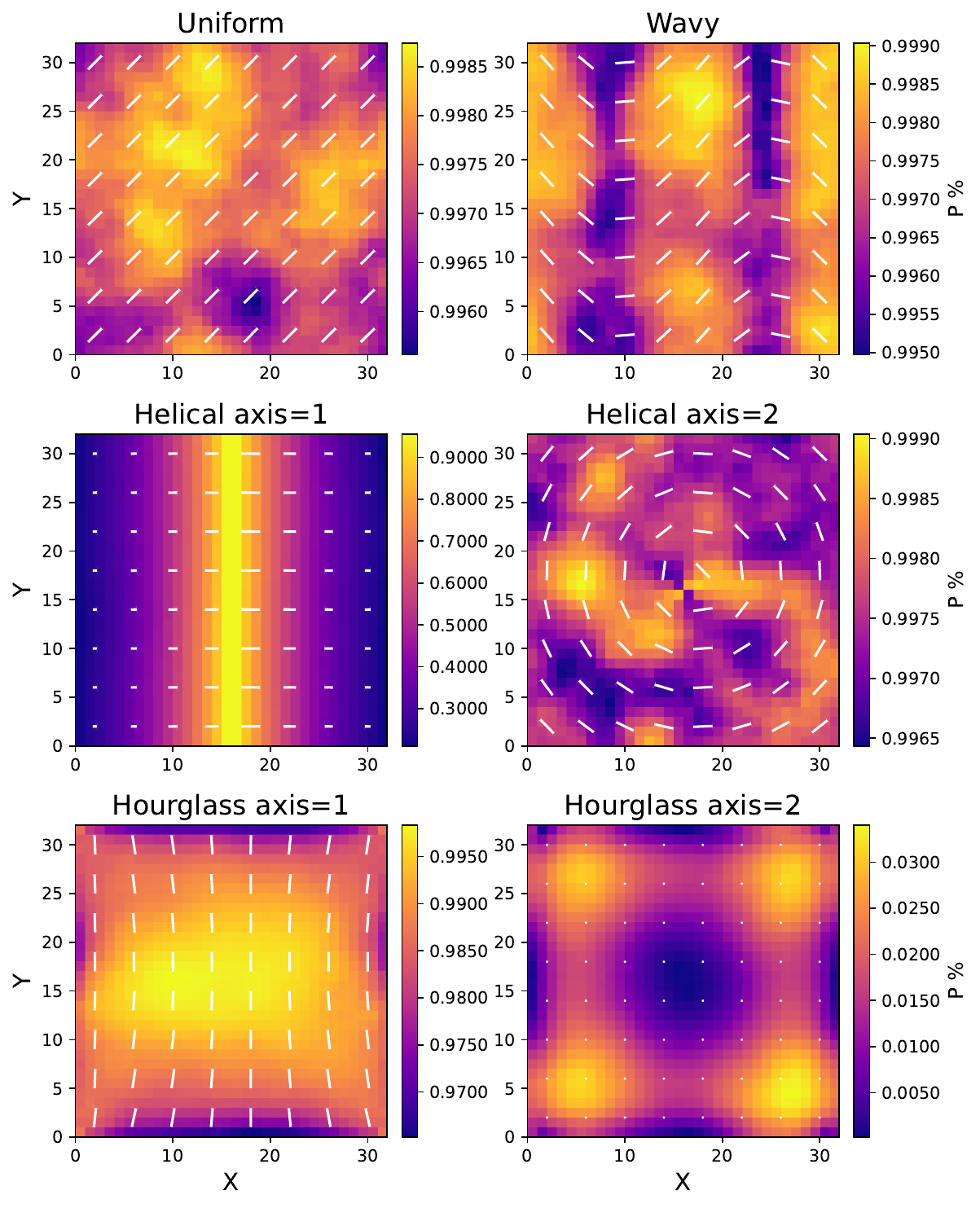}
\caption{Synthetic polarization maps for Uniform, Wavy, Helical, and Hourglass morphologies. Polarization fraction is shown as a heatmap. Here, models show M$_A = 0.01$ and $\sigma_{Q,U}$ to highlight the underlying geometric templates.}
\label{fig:pol_maps}
\end{figure}

Figure~\ref{fig:pol_maps} shows the synthetic PA map overlaid on the polarization fraction for the geometric templates tested. 
Regions of depolarization due purely to geometric effects are apparent in the Hourglass template, while Uniform, Wavy, and Helical are mostly in the POS and fully polarized.

\section{Results}
\label{sec:results}

To what extent can the magnetic field geometry be recovered, given only polarization information?
We propose a decision tree classifier based solely on polarization statistics to recover magnetic field geometry. 
Figure~\ref{fig:decision_tree} and Section~\ref{sec:decisiontree} outline the decision tree, measurements, and results. 
The following details each polarization statistic, or measurement, used in the classifier.

\subsection{P and PA Distributions}
\begin{figure}[t]
\centering
\includegraphics[width=0.38\paperwidth]{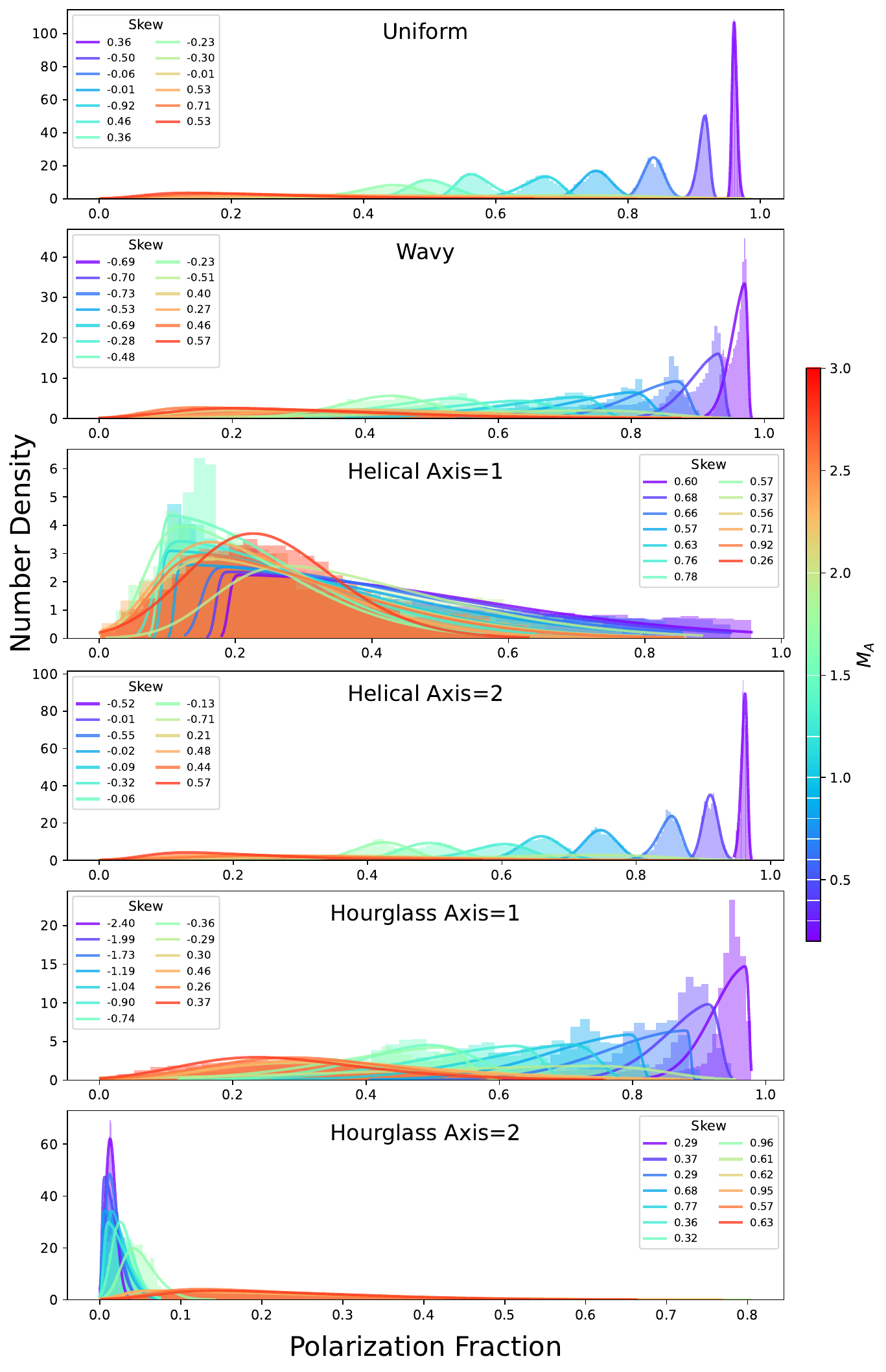}
\caption{Polarization fraction distributions at varying ${\cal{M}}_A$ numbers at $\sigma_{Q,U}=0$. The distributions shift to lower polarization fraction as turbulence increases in all cases but the Hourglass Axis = 2 and Helical Axis = 1, which already experience high levels of depolarization from geometric and LOS integration effects.}
\label{fig:pol_dists}
\end{figure}

Figure~\ref{fig:pol_dists} shows the polarization fraction distribution, for each geometry, for a range of ${\cal{M}_A}$ values, $0.1\leq {\cal{M}_A}\leq3.0$.
Specifically, we measure the P distribution peak and median as statistical measures of the characteristic distribution for each geometry (see Fig.~\ref{fig:decision_tree}).

For sub-\alf cases (Fig.~\ref{fig:pol_dists}, purple), Uniform, Wavy, Helical Axis=2, and Hourglass Axis=1 share distribution shape and mean. 
These fields are primarily in the POS and near 100\% polarization fraction. 
For sub-\nalf, any depolarization will occur mainly from geometry and/or LOS integration. 
This is the effect seen for Helical Axis = 1 and Hourglass Axis = 2, where the P distribution is closer to 0-30\%.
The field in Helical Axis = 1 is depolarized from overlapping field, highlighting depolarization from LOS integration.
Meanwhile, the magnetic field in Hourglass Axis = 2 is mostly in the LOS, showing depolarization due to geometric shape.
These structural differences account for the low P seen in Figure~\ref{fig:pol_dists} for sub-\alf cases. 

For super-\alf cases (Fig.~\ref{fig:pol_dists}, red), the P distribution broadens and the overall fraction decreases for most geometries due to turbulence.
The field becomes essentially random/chaotic.
Seen in Fig.~\ref{fig:pol_dists}, the Uniform, Wavy, Helical Axis = 2, and Hourglass Axis = 1 fields experience a left shift (decrease) in the P distribution as \sma increases. 
Helical Axis = 1 and Hourglass Axis = 2 shift right (increases), and turbulence effectively adds polarization.
Above \ma$\sim2$, all geometries converge to the same P distribution -- a random field.

\begin{figure*}[t]
\centering
\includegraphics[width=0.75\paperwidth]{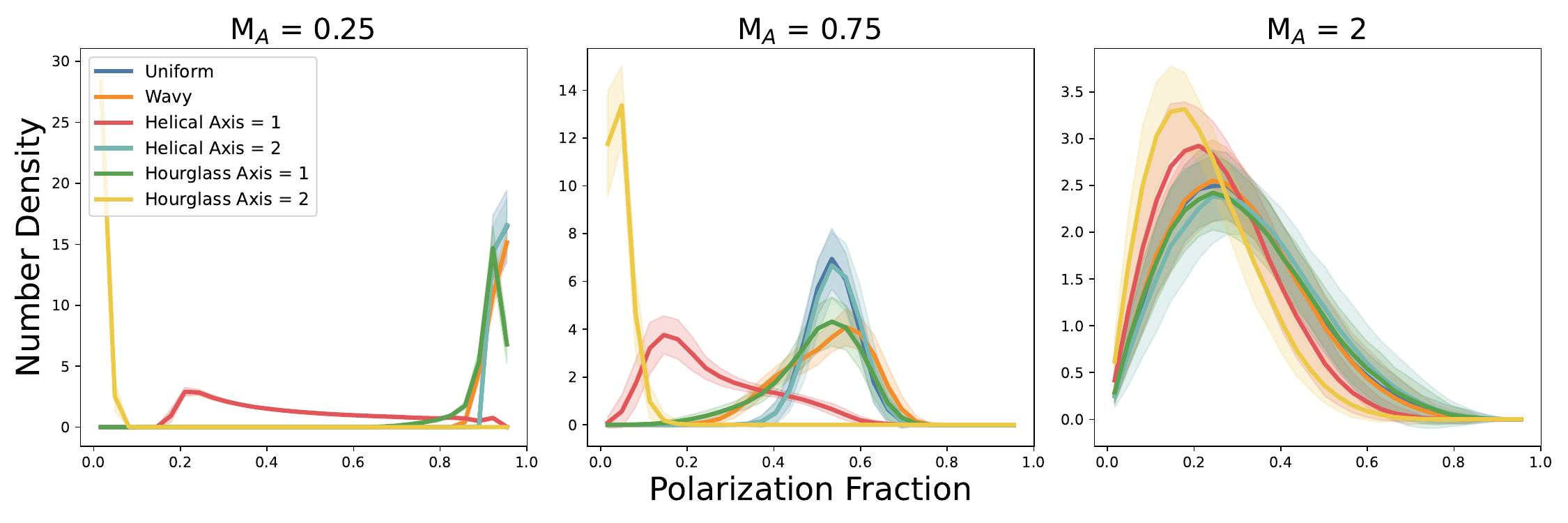}
\hfill
\includegraphics[width=0.75\paperwidth]{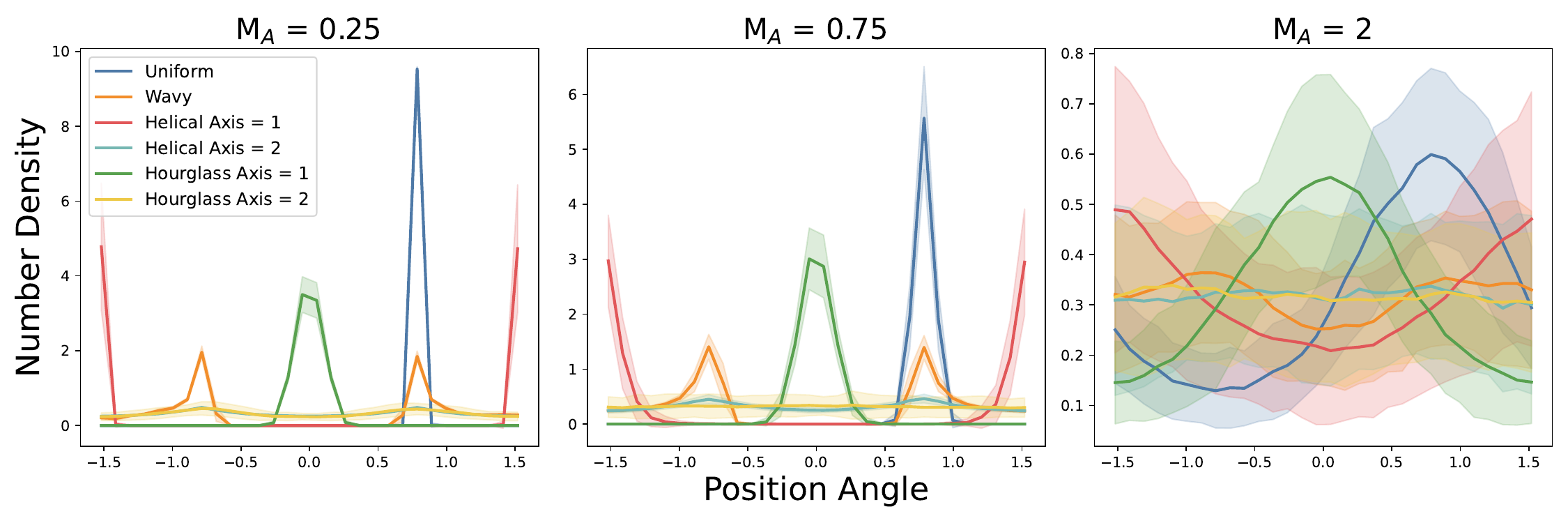}
\caption{Average P and PA distributions with 1$\sigma$ uncertainty for 100 runs for various geometries. Distributions here are shown for M$_A = [0.25, 0.75, 2]$. The greatest difference in distribution across geometries is at M$_A =0.25$, when magnetic field geometry is most prominent. At M$_A =2$, turbulence dominates and all geometries converge towards the same distribution.}
\label{fig:dist_compare}
\end{figure*}

Figure~\ref{fig:dist_compare} compares every template at three different \sma values [0.25, 0.75, 2] to test whether geometries are distinct in polarization (P) and position angle (PA) distribution. 
The mean and 1-sigma uncertainty are derived from 100 independent realizations of the experiment.
\sma values were selected to represent the sub-, trans-, and super-\alf environments.

For sub-/trans-\alf cases (\ma$=0.25,0.75$) magnetic field geometry is the strongest and we see the most distinct P distributions (Fig.~\ref{fig:dist_compare}, first row). 
Hourglass Axis = 2 and Helical Axis = 1 show unique distributions, with low P resulting from geometry and LOS integration.
Meanwhile, the Uniform, Wavy, Helical Axis = 2, and Hourglass Axis = 1 are all similar in polarization distribution, and not distinguishable. 
At \ma$=2$, distributions and geometries are indistinguishable as turbulence saturates. 

The PA distribution (Fig.~\ref{fig:dist_compare}, second row) is also a useful statistic in identifying geometry and turbulence regime.
However, it should be noted that the mean of the distribution can vary depending on the viewing angle of the field, so only shape should be used to separate geometries. 
For sub/trans-\alf cases (\ma$=0.25,0.75$), most geometries are somewhat distinguishable, with some degeneracies. 
Uniform shows a saturated peak, which can be degenerate with Hourglass Axis = 1 for a given viewing angle. 
Hourglass Axis = 2 and Helical Axis = 2 share the same broad distribution with no clear peaks, and are therefore degenerate. 
Wavy stands alone as a double or multi-peak distribution (depending on $k$ or scale of observation).
Helical Axis = 1 is also unique in having two distinct populations at high and low PA ($\sim \pm 90^{\circ}$).

For super-\alf cases, turbulence once again saturates and the PA becomes too noisy to differentiate geometry, as seen by the 1-sigma uncertainty as the shaded regions in Figure~\ref{fig:dist_compare}.

\subsection{Polarization Distribution Skewness}

\begin{figure}[t]
\centering
\includegraphics[width=0.38\paperwidth]{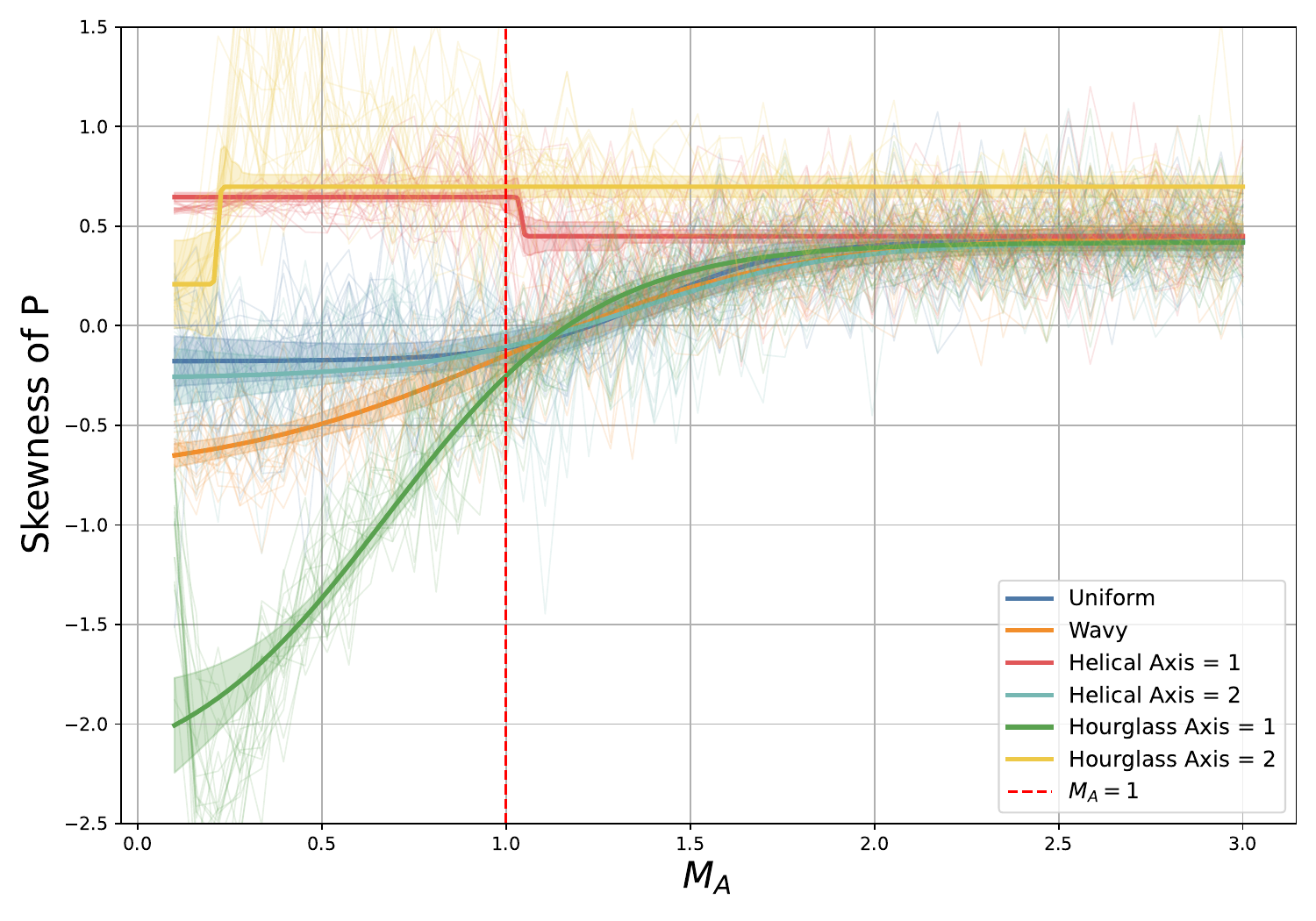}
\caption{Polarization fraction distribution skewness as a function of ${\cal{M}}_A$. Data is fitted with a $f_{tanh}(x)$ model. 1$\sigma$ uncertainty is plotted as the shaded regions around each fit. As the ${\cal{M}}_A$ increases, Kolmogorov turbulence dominates and all geometries converge on the same skewness. A red dashed line is set at M$_A = 1$ to highlight the difference in skewness values between sub/trans- and super-\nalf. Skewness may be a helpful metric in establishing the level of turbulence.}
\label{fig:skewness_trend}
\end{figure}

At each ${\cal{M}}_A$, we also find the polarization distribution skewness.
Figure~\ref{fig:skewness_trend} shows the skewness of the polarization fraction distribution for each geometry as a function of ${\cal{M}}_A$. 
To fit the data, which is seen as faint histograms, we used the Akaike Information Criterion (AIC; Akaike 1974) for model selection,
\begin{equation}
    AIC = n \ln\frac{SSE}{n} +2k
\end{equation}
where n is the number of data points, k is the number of free parameters, and SSE is the sum of squared errors. 
AIC penalizes additional free parameters and allows comparison across models.
The following models were used,
\begin{align}
    f_{tanh}(x) &= a*\tanh(bx + c) + d \\
    f_{rational}(x) &= \frac{a-b}{x^c+d} \\
    f_{reciprocal}(x) &= \frac{a-b}{x}\exp(-cx).\\
\end{align}
It was found that a $f_{tanh}(x)$ returned the lowest AIC value, which we accepted as best fit.
The $f_{tanh}(x)$ model was then fit using nonlinear least squares regression as implemented in \texttt{scipy.optimize.curve\_fit} (Virtanen et al. 2020).

Figure~\ref{fig:skewness_trend} shows each geometry with a $f_{tanh}(x)$ fit for 50 runs, with 1$\sigma$ uncertainty plotted as the shaded region. 
The red dashed line marks the transition between sub/trans-\alf and super-\alf (\ma$=1$). 
At \ma$=2$, all geometries approach the same skewness as turbulence dominates and approaches Kolmogorov.

For sub/trans-\alf cases, the skewness is highly dependent on the geometry. 
The Uniform, Wavy, and Helical Axis = 2 fields are almost entirely in the POS and all possess more uniform distributions, which means a less steep skewness fit, as seen in Figure~\ref{fig:dist_compare}.
Hourglass Axis = 1 feels some depolarization at the edges, which creates a tail in P distribution. 
Thus, the skewness trend is steeper and more distinct.
Helical Axis = 1 and Hourglass Axis = 2 are highly depolarized and so their distribution skewness begins higher. 
For Helical Axis = 1, the addition of turbulence actually increases polarization fraction and decreases the skewness.
Based on these geometric characteristics, the polarization skewness can provide a metric of whether the environment is sub/trans- or super-\nalf.

\subsection{PA Circular Variance}

\begin{figure}[t]
\centering
\includegraphics[width=0.38\paperwidth]{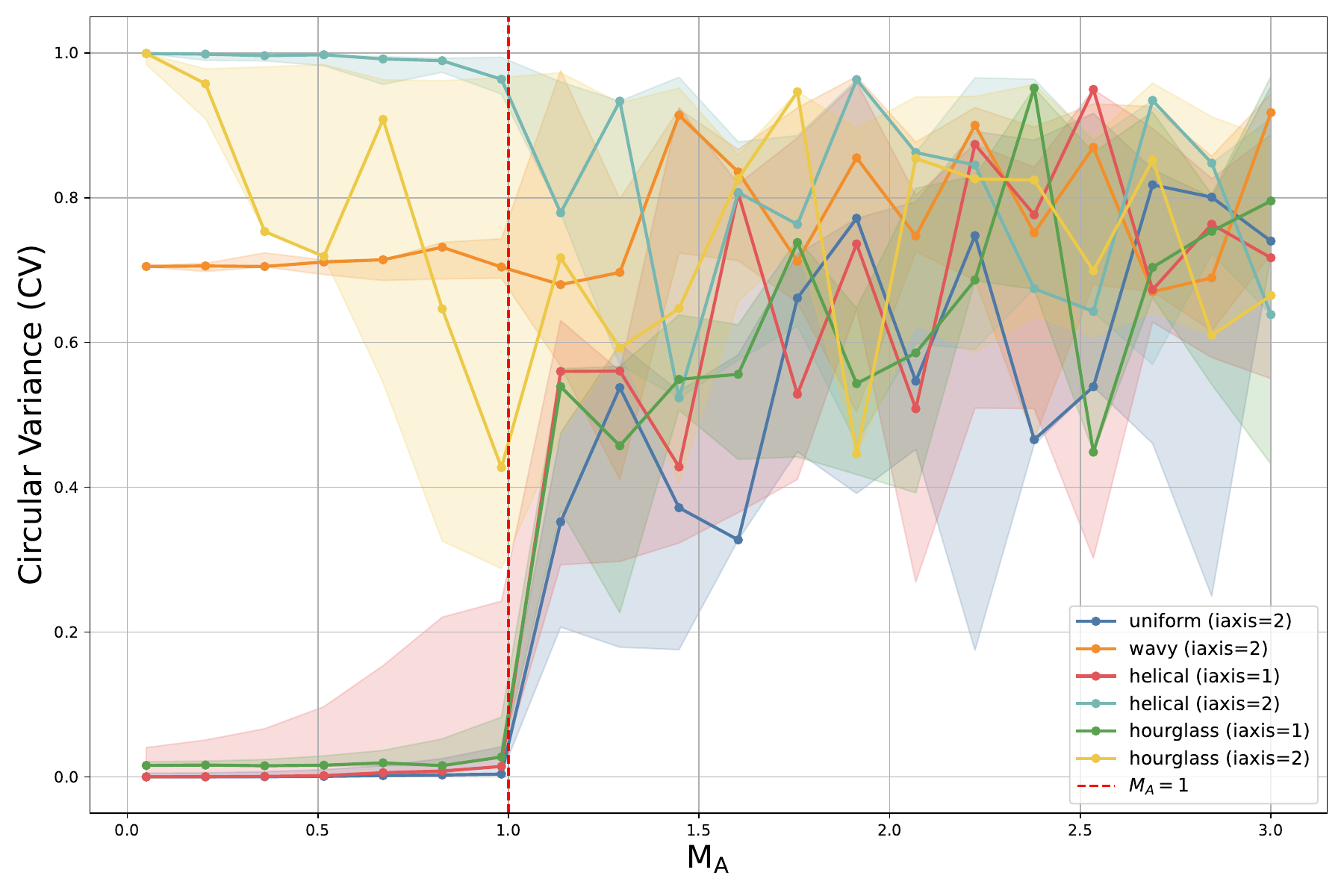}
\caption{Circular Variance of position angle (PA) at 10\% injected noise to highlight its ability to distinguish geometries. For sub-\alf cases, two distinct populations emerge, at CV$_{\rm PA}<0.1$ and CV$_{\rm PA}>0.65$. This statistic provides an additional way to break degeneracies.}
\label{fig:CV}
\end{figure}

Quantifying the PA distribution can also be used to classify geometry and turbulence. 
Following \citep{Shane+2024}, the PA dispersion, $\bar S$, is calculated. 
Similar dispersion values were found that matched \citet{Shane+2024} for sub- and super-\alf environments.
However, after the injection of 10\% noise, $\bar S$ failed to distinguish between geometries. 
Thus, we implemented the circular variance (CV$_{\rm PA}$) of the PA, which is calculated by,
\begin{align}
    \rm \bar R &= | \frac{1}{N} \sum_{i=1}^N e^{2i\phi_i}| \\
    \rm CV_{PA} &= 1 - \bar R
\end{align}
where $\bar R \in [0,1]$ is the mean resultant length, N is the number of PA measurements, and $\phi_i$ is a single PA measurement.
CV$_{\rm PA}$ $\in[0,1]$, where CV$_{\rm PA}$=0 is a perfectly ordered field and CV$_{\rm PA}$=1 is a fully random field. 
PA is circular in definition, where $180^\circ = 0^\circ$, so this CV$_{\rm PA}$ metric is particularly useful.

The CV$_{\rm PA}$ is found as a global mean over all measurements and suppresses small scale noise. 
This in turn means smaller variations in the CV$_{\rm PA}$ value across noise injections -- and therefore better for identifying geometries. 

Figure~\ref{fig:CV} shows the CV$_{\rm PA}$ at $\sigma = 10\%$ across ${\cal{M}}_A$ values for each geometry. 
Histograms represent 50 runs averaged, similar to Figure~\ref{fig:dist_compare}.
1$\sigma$ uncertainty is shown as the shaded region. 
There is a clear distinction between sub/trans- and super-\alf environments, separated by the red dashed line.

There appear to be two populations of CV$_{\rm PA}$ values for sub-\alf cases: Helical Axis = 1, Uniform, and Hourglass Axis = 1 at CV$_{\rm PA}<0.1$, and Wavy, Helical Axis = 2, Hourglass Axis = 2 at CV$_{\rm PA}>0.65$. 
This provides another way to break degeneracies.
We note that the first group can serve as a way of classifying the ${\cal{M}}_A$ regime, the second group, not so much.

\subsection{Median Polarization Fraction}

\begin{figure}[t]
\centering
\includegraphics[width=0.38\paperwidth]{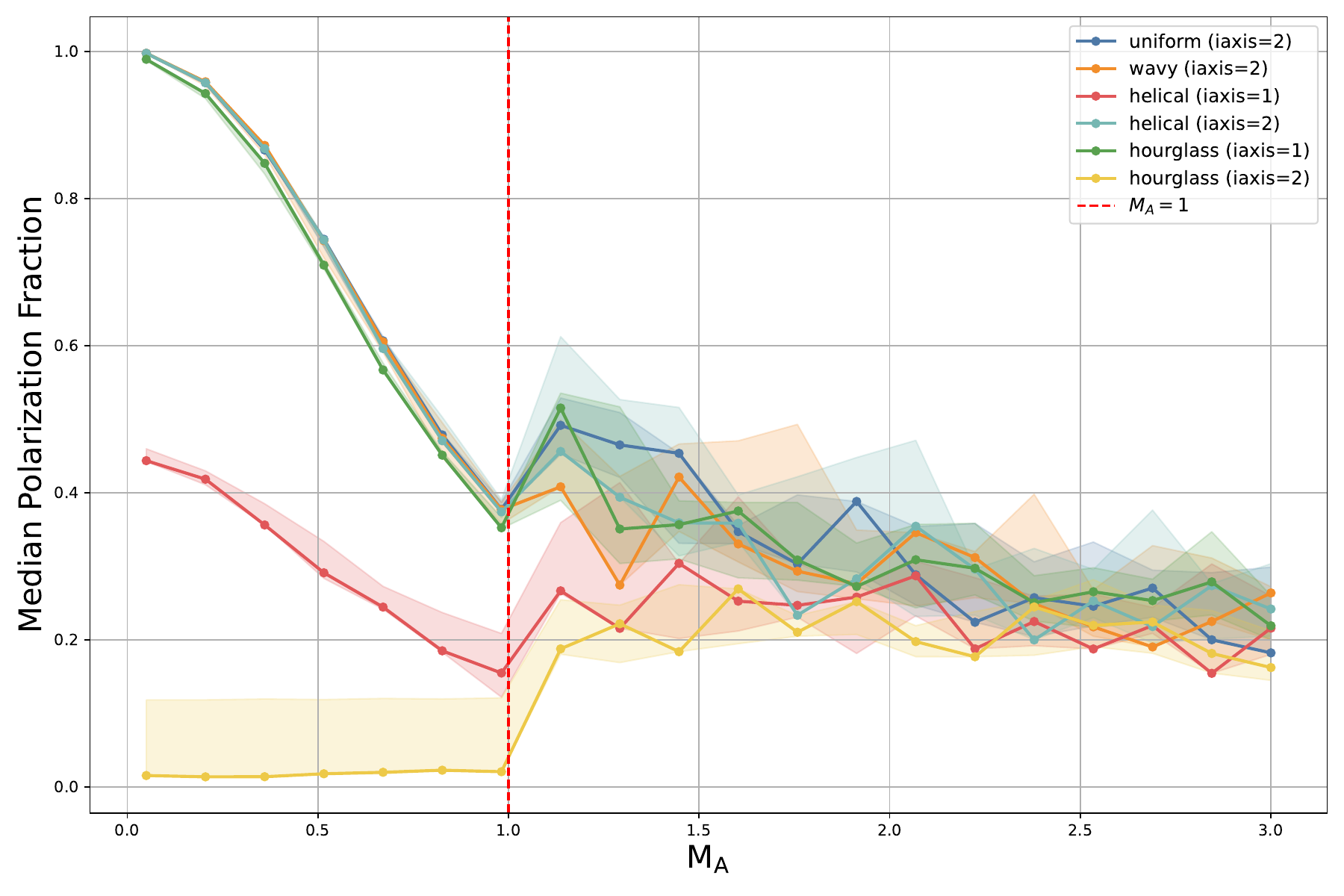}
\caption{Median Polarization fraction for each geometry is shown for varying M$_A$ at 10\% injected noise. The average histogram after n=50 runs is shown as the solid line, with 1$\sigma$ uncertainty as the shaded region.}
\label{fig:med_P}
\end{figure}

The median polarization fraction ($\rm \bar P$) is a straightforward yet powerful statistic -- turbulence, once super-\nalf, heavily influences P, while geometry is the main informer for sub-\alf cases.
Figure~\ref{fig:med_P} shows these effects. 
While the $\rm \bar P$ may not be the best at classifying the $\cal{M}_{\rm A}$ environment, for sub-\alf cases the geometries exist in clear groupings.
Hourglass Axis = 2 and Helical Axis = 1 are completely unique in $\rm \bar P$ for sub-\alf cases. 
The rest of the geometries are indistinguishable from each other, but sit at much higher values of P in sub-\alf than in super-\alf environments generally.

\section{Decision Tree}
\label{sec:decisiontree}

\begin{figure*}[t]
\centering
\includegraphics[width=0.83\paperwidth]{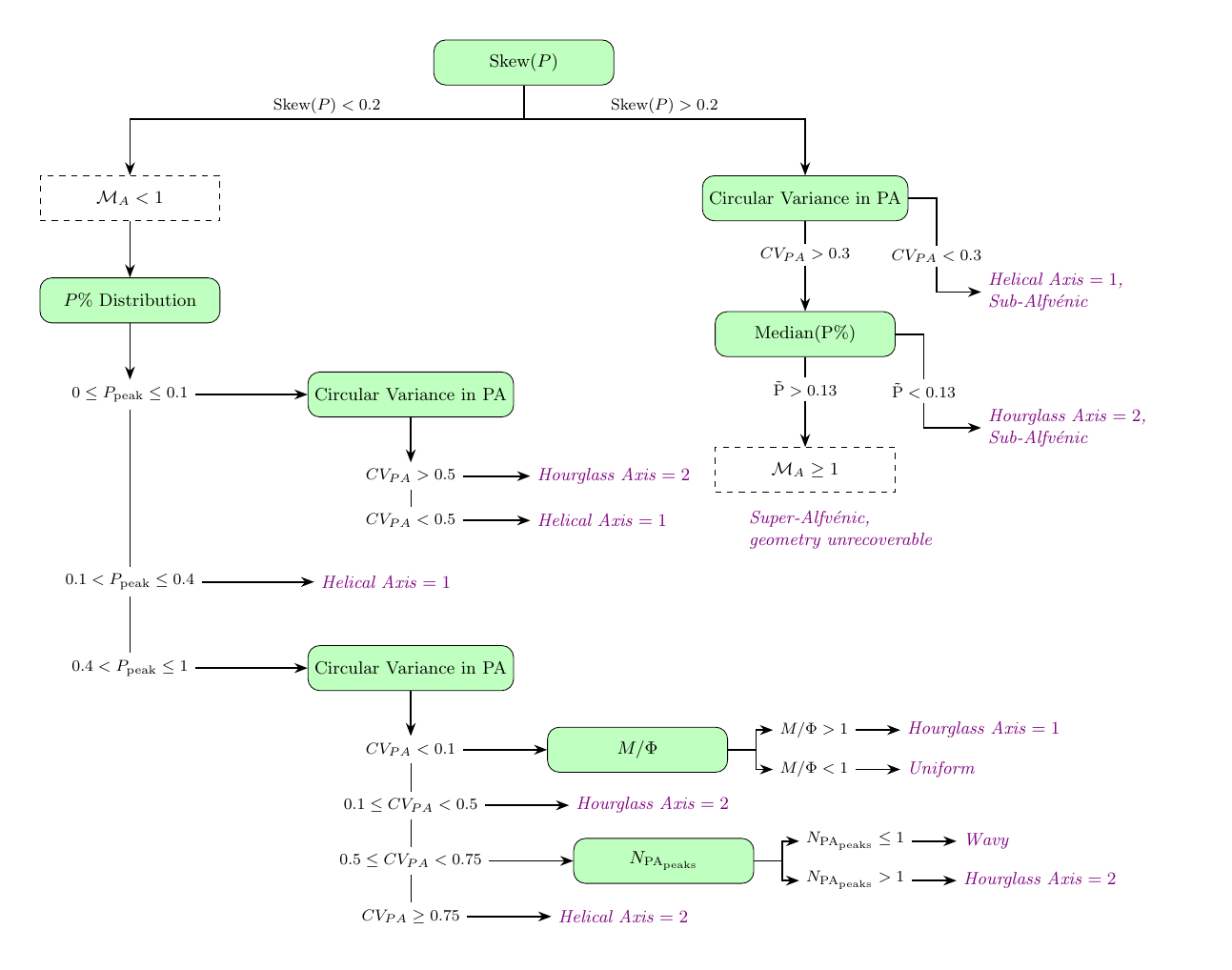}
\caption{Similar to \citet{Shane+2024}, a decision tree can help determine the magnetic field geometry of the observed field. Here, we generate a logic based on a series of measured synthetic polarization statistics.}
\label{fig:decision_tree}
\end{figure*}

As seen in previous sections, there exists some degeneracy in every polarization statistic, which is why measuring the magnetic field geometry from polarization alone has been so difficult.
However, with a combination of measures, we can form a classification method that breaks these degeneracies.

Figure~\ref{fig:decision_tree} outlines the decision tree, with green modules indicating a measurement and magenta lettering the result.
The numerical values for each decision were determined for 50 realizations, across noise injection levels, for varying \sma values, as described in the previous Section~\ref{sec:results}.

Beginning at the top of the tree in Figure~\ref{fig:decision_tree}, we use the skewness in P to determine whether the environment is sub/trans-\alf or super-Alfv\'{e}nic.

There is degeneracy for skew(P) $> 0.2$; the field could either be super-\nalf, Hourglass Axis = 2, or Helical Axis = 1.
Thus, we use an additional measure to break this: CV$_{\rm PA}$ $< 0.3$ indicates a Helical Axis = 1 field in sub-\alf; CV$_{\rm PA}$ $> 0.3$ and $\rm \bar P<0.13$ indicates a Hourglass Axis = 2 and sub-\alf field; else the field is super-\nalf. 

If skew(P)$<0.2$, the environment is determined to be sub/trans-\nalf. 
From here, the geometry can be determined through a series of measurements: the peak of the P distribution (P$_{peak}$), CV$_{\rm PA}$, mass-to-flux Ratio ($M/\Phi)$, and number of peaks in the PA distribution ($\rm N_{PA_{peaks}}$).

Measurements near the end of classification were specifically implemented for the remaining degenerate geometries. 

The mass-to-flux ratio was used to differentiate between Hourglass Axis = 1 and Uniform fields, since theoretically they exist in very different environments. 
Computing the mass-to-flux ratio requires a column density; we adopt $N_{\rm H_{2}}$ since dust-based tracers have been shown to give more reliable column density estimates in star-forming clouds rather than CO-based tracers \citep{Goodman+2009}.
However, our simulations are scale-free, so we use previous observational findings of magnetic field geometries to determine column density: Uniform $\sim10^{20-21} \rm cm^{-2}$ \citet{Planck+2015} and
Hourglass $\sim10^{22}-10^{23} \ \rm cm^{-2}$ \citet{Girart+2006, Qiu+2014}. 

The number of peaks in the PA distribution ($\rm N_{PA_{peaks}}$) was used to distinguish between Wavy and Hourglass Axis = 2 fields. Looking at Figure~\ref{fig:dist_compare}, the Wavy field will contain distinct peaks depending on the frequency of the Wavy. 
Hourglass Axis = 2 PA values are circular and will always show a broad PA distribution. 
Thus, the number of peaks in the PA distribution was an effective way to break this degeneracy.

\begin{figure*}[t]
\centering
\includegraphics[width=0.8\paperwidth]{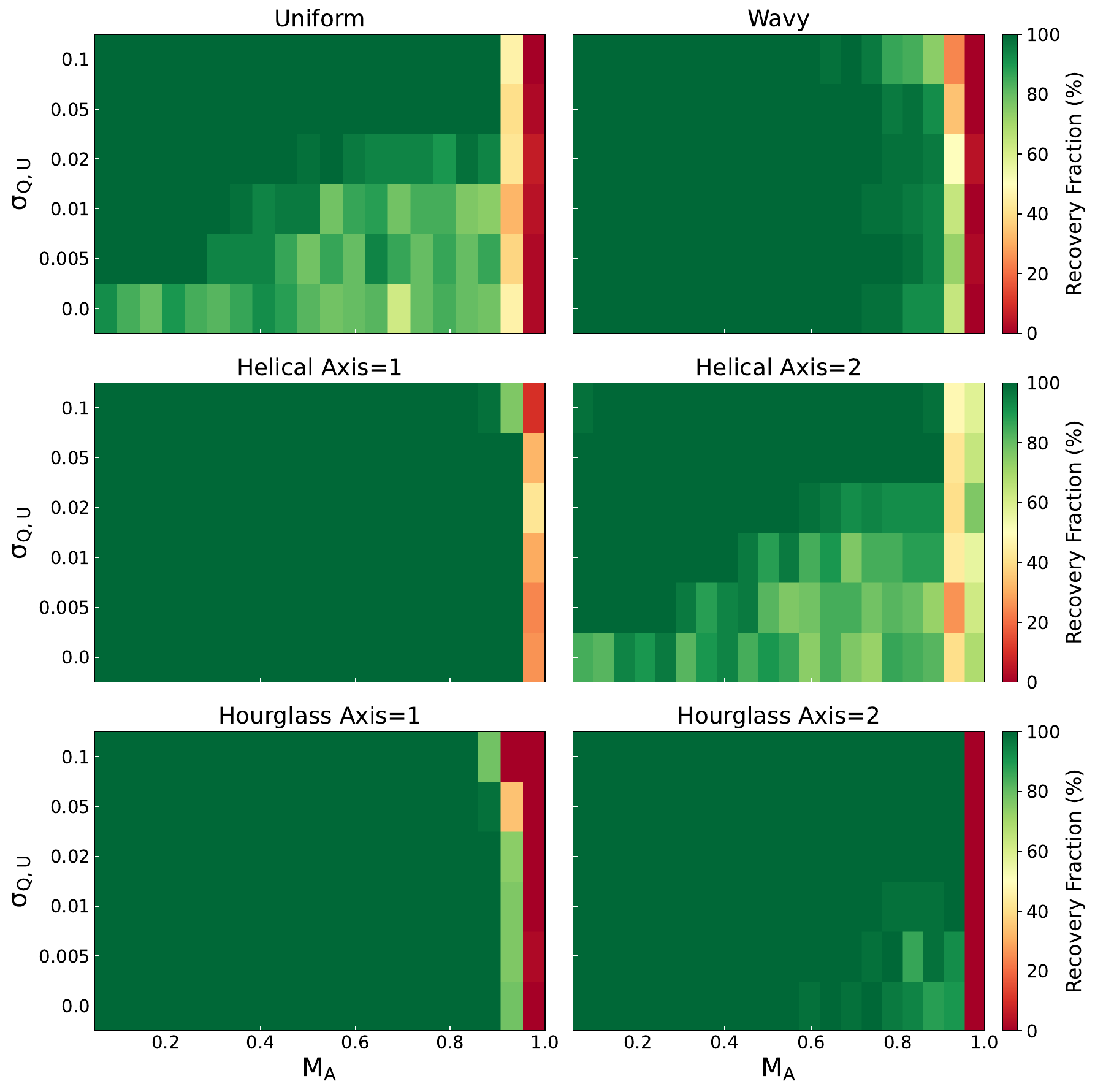}
\caption{Recovery percentage heat maps as a function of injected Stokes noise (y-axis, $\sigma_{Q,U}$) and \alf Mach number (x-axis, \ma) for 50 realizations. Recovery systematically fails as \ma $\xrightarrow{}1$ and turbulence saturates the field. All geometries are recovered well ($>80\%$) up to \ma$=1$, with the exception of Uniform and Helical Axis = 2, which struggle to be identified due to their low skew(P) values for sub-\alf cases.}
\label{fig:recovery}
\end{figure*}

We quantify the effectiveness of the decision tree by running 50 realizations at a series of noise injections $\sigma_{Q,U} = [0,10\%]$, across sub-\alf values $\rm M_A = [0,1]$. 
Figure~\ref{fig:recovery} shows the recovery fraction percentage for each geometry.
The decision tree appears to break down globally as it approaches \ma$\xrightarrow{}1$. 

Uniform and Helical Axis = 2 have lower likelihoods of being classified correctly due to the first decision in Fig.~\ref{fig:decision_tree}: the skew(P).
The skew trends for Uniform and Helical Axis = 2 (see Fig.~\ref{fig:skewness_trend}) are not steep and hover around 0, making the decision skew(P)$<0.2$ difficult to achieve.
Interestingly, the addition of noise increases the likelihood of correct classification for the two geometries.
We suspect the addition of noise broadens and therefore skews the distribution, thus making the skew trend (seen in Fig.~\ref{fig:skewness_trend}) steeper and recovery easier. 

\section{Discussion}
\label{sec:diss}
Recovering the 3D magnetic field geometry is not straightforward. 
Resolving the LOS component from observations is indirect and degenerate. 
As mentioned previously, the observed polarization signal entangles contributions from turbulence, magnetic field geometry, and LOS integration effects.
This study was formulated to account for all three. 
An incompressible, anisotropic turbulence model was prescribed to perturb the magnetic field.
Magnetic field geometries with configurations primarily aligned along the LOS (e.g., Hourglass Axis = 2) were implemented to address depolarization arising from field direction. 
Finally, we included the Helical geometry to capture LOS integration effects, where depolarization results from cancellation.

Our models (i.e., scale-free and time-independent) are highly idealized compared to full MHD simulations \citep{Stone+2020, Ostriker+2001, Burkhart+2020}.
This idealization allows for flexibility in geometry selection, which would otherwise be difficult to achieve.  
Although geometric templates can limit the method's applicability, we restrict ourselves to templates that have been previously discussed or found in the ISM, similar to the template selection in \citet{Reissl+2021}. 
However, our approach differs from \citet{Reissl+2021} in its goal: they apply forward-modeling to produce synthetic observables for comparison to individual observations, whereas we use forward-modeling to derive polarization statistics that inform a decision tree classifier, enabling systematic geometry recovery.

For completeness, we compared our results against MHD simulations from the Catalogue for Astrophysical Turbulence Simulations (CATS) \citep{Cho+2003, Burkhart+2009, Portillo+2018, Bialy+2020, Burkhart+2020}.
We used 3D driven, isothermal, compressible MHD turbulence (Cho-ENO) models at two Alfv\'{e}n Mach numbers. 
The simulations begin with a uniform field that subsequently evolves under the driven turbulence. 
We compared the polarization fraction distribution and its skewness at \ma = 0.7 and 2 (sub- and super-\nalf, respectively; see Appendix~\ref{appendix} for details). 
Agreement is dependent on environment: our GS95 prescription matches the MHD skewness in the sub-\alf case, but deviates in the super-\alf case, likely due to our treatment of anisotropy relative to a single global field direction rather than a spatially-varying local one.
Although there are apparent discrepancies, they agree in the overall trend of polarization skewness -- as \ma increases, distributions shift from right to left, and skewness becomes more positive.

\citet{Shane+2024} use the Zeeman effect to measure the LOS Alfv\'en Mach number (${\cal{M}}_{A,z}$) and show that \ma can be estimated from the dispersion in position angle (S) via a decision tree. 
Our method can complement this approach in cases where the skew(P) decision tree fails to reliably identify the turbulence regime.
Alternatively, when Zeeman information is available, \citet{Shane+2024}'s method can be combined with our decision tree to yield a more accurate joint prediction of cloud characteristics and geometry.

Our classification method performs well for sub-\alf cases but breaks down in for super-\alf, where all polarization statistics become degenerate. 
For super-\alf environments, a cloud is dominated by turbulence and polarization statistics lose their distinct, geometry-dependent signatures.
Although geometric recovery via classifier is not possible in this case, in these environments the field is passively advected by the flow -- its geometry can then be inferred directly from the velocity field rather than from polarization statistics.
For sub-\alf cases, by contrast, the cloud is likely magnetically regulated, in regions where star formation is just beginning.

How well this method holds up across varying observing angles remains an open question. 
Viewing the Hourglass and Helical geometries along different axes shows that the inferred field structure changes substantially with orientation.
Many star-forming filaments are known to exist at some inclination angle, as derived from velocity profiles (e.g., Orion A; \citealt{Stutz+2016, Grobschedl+2018}).

While the classification system performs well in model space, a major caveat is the absence of grain alignment effects. 
The models assume 100\% grain alignment and efficiency and constant density.
In reality, imperfect grain alignment and efficiency suppress the polarization fraction, shifting the P distribution toward lower values.
Hence, this is a first step, demonstrating that the method may work in principle. 
As a next step, we plan to incorporate density and grain alignment and efficiency effects into the models.

The method presented here is a basis for unraveling the LOS component in common geometries, and it can be extended by incorporating additional geometric templates as they are identified.

\section{Summary}
\label{sec:sum}
We develop a decision tree classifier based on polarization statistics that distinguish between 3D magnetic field geometries.
We test the classifier on four geometric templates, Uniform, Wavy, Helical, and Hourglass, including multiple viewing angles for the asymmetric Helical and Hourglass geometries. 
We perturb the magnetic field with anisotropic, incompressible turbulence following \citet{Goldreich+1995} and \citet{Cho+2003}, spanning varying \alf Mach numbers (\ma) for sub-, trans-, and super-\alf environments. 

For each realization, we measure a suite of polarization statistics: the peak (P$_{\rm peak}$), median ($\rm \bar P$), and skewness of the polarization fraction distribution (skew(P)); the circular variance of the position angle (CV$_{\rm PA}$); the number of position angle peaks ($\rm N_{PA_{peaks}}$); and the mass-to-flux ratio ($M/\Phi$).
We find that each individual statistic is degenerate with respect to geometry and no single statistic uniquely identifies the underlying field structure.
To resolve these degeneracies, we combine these statistics into a decision tree classification scheme and test it on a series of \alf Mach numbers \ma$\in[0.1,1]$ and injected Stokes noise $\sigma_{Q,U} \in [0,10\%]$.
The classifier can differentiate between the geometries well on average ($>80\%$), with the exception of Uniform and Helical Axis = 2, which struggle to overcome the first branch in the decision tree at the skew(P) measurement.

This framework offers a starting point for disentangling LOS effects across common field geometries, and is readily extensible to additional templates as they are identified. 
As an idealized first step, incorporating density variations and grain alignment efficiency is the next extension of this work.

\section{Acknowledgments}
This work was supported by the NASA Future Investigators in NASA Earth and Space Science and Technology (FINESST) award, proposal number 23-ASTRO23-0053.

The simulation and decision tree classification code used in this study are available upon request. 
The CATS MHD simulation data used for comparison in Appendix~\ref{appendix} are publicly available \citep{Burkhart+2020}.

\appendix

\section{Comparison to MHD, Kolmogorov Models}
\label{appendix}

\begin{figure*}[t]
\centering
\includegraphics[width=0.48\textwidth]{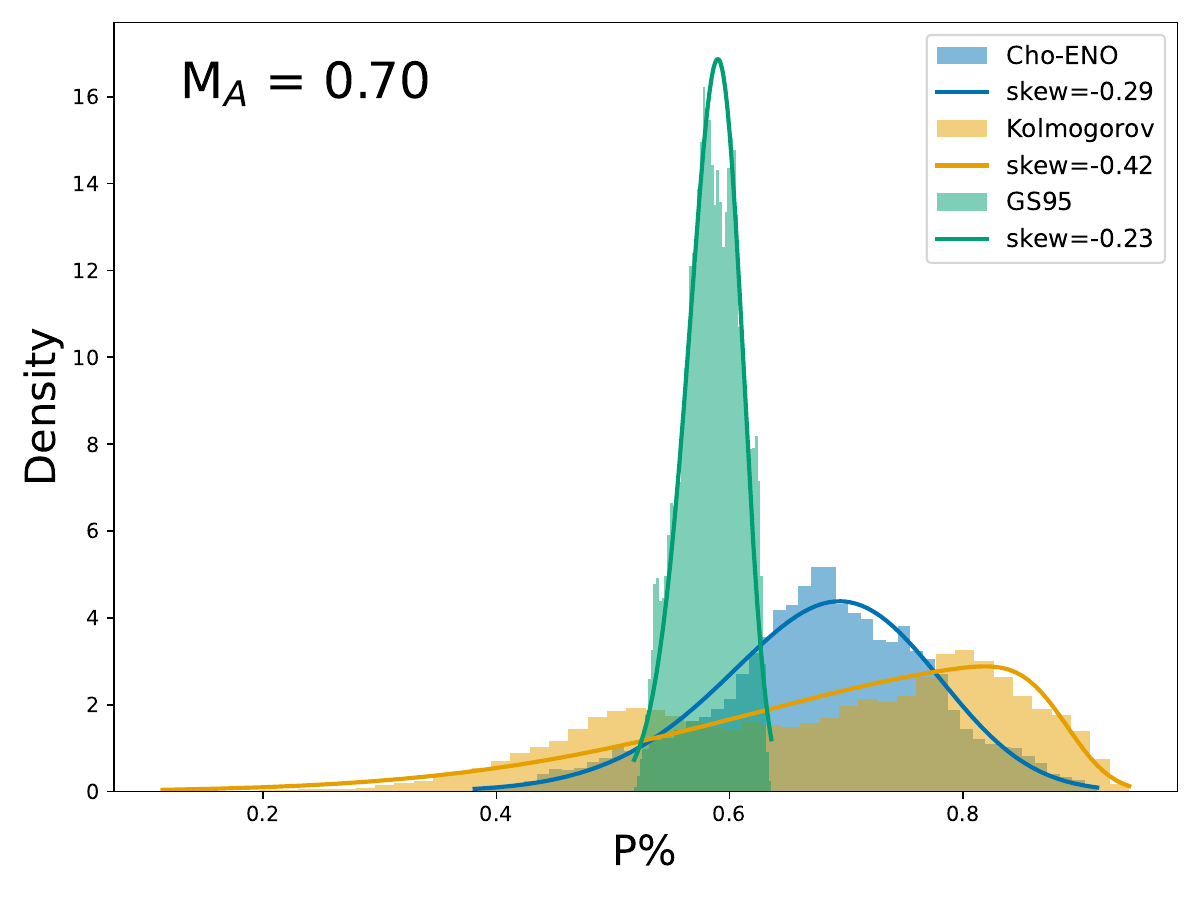}
\hfill
\includegraphics[width=0.48\textwidth]{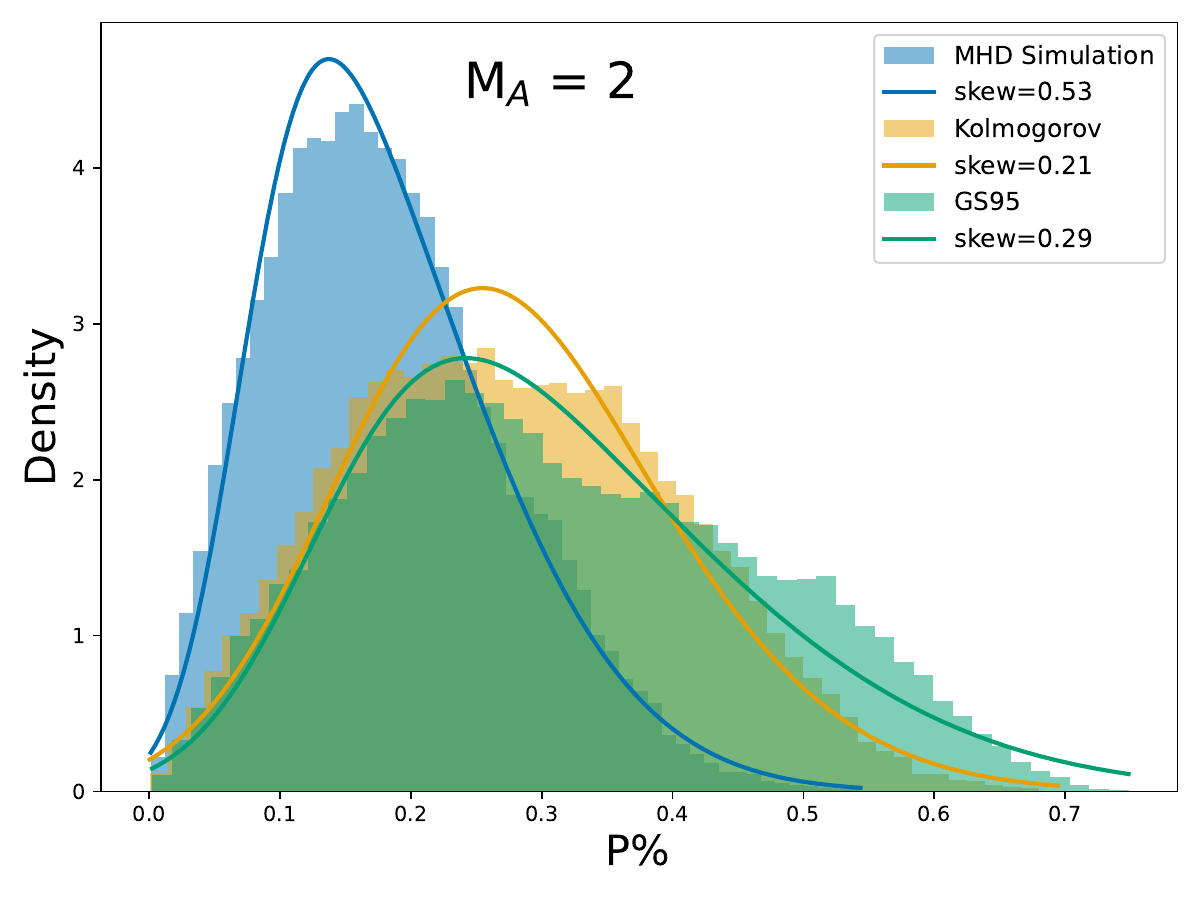}
\caption{Polarization fraction distributions for Cho-ENO (CATS), Kolmogorov, and GS95 turbulence models for sub-\alf (\ma=0.7, left) and super-\alf (\ma=2, right). 
We find that agreement is dependent on the turbulence environment.
At \ma=0.7, the GS95 model matches the Cho-ENO closely in skewness, albeit their distribution shapes vary greatly.
At \ma=2, the Kolmogorov model now matches GS95's skewness and shape, Cho-ENO showing a greater skew.
We find that these discrepancies, in distribution shape and skewness, are a result of our global anisotropic prescription.
However, all three models show a global trend of increasing polarization distribution skewness for increasing \alf Mach number.}
\label{append:MHD_compare}
\end{figure*}

To test the effectiveness of our turbulent magnetized model, we can compare to the Catalogue for Astrophysical Turbulence Simulations (CATS) \citep{Burkhart+2020}.
Figure~\ref{append:MHD_compare} compares the polarization fraction skewness of our model against the Cho-ENO simulations, as well as a pure Kolmogorov model, for sub-\alf (\ma$=0.7$) and super-\alf (\ma$=2$) cases at equal resolutions (256x256x256). 
Cho-ENO simulations are 3D numerical experiments of compressible (MHD) turbulence and use a third-order-accurate hybrid essentially nonoscillatory (ENO) scheme to solve the ideal MHD equations in a periodic box \citep{Cho+2003, Burkhart+2009, Portillo+2018, Bialy+2020, Burkhart+2020}.

At \ma$=0.7$, the Cho-ENO simulation's skewness (-0.29) is matched closely by the GS95-based prescription (-0.23), while the Kolmogorov model (-0.42) deviates. We also note the overall concentrated shape of the GS95-based model when compared to the Cho-ENO or Kolmogorov model.

At \ma$=2$, the skewness of the Kolmogorov model (+0.21) is closer to the GS95-based model (+0.29) than the skewness of the Cho-ENO simulation (+0.53). At this point, Kolmogorov turbulence is dominating in the GS95 model, explaining the similar distribution shapes the two models share. Cho-ENO instead feels more depolarization and a higher skew. 

These slight discrepancies in skew and distribution shape likely reflect a structural limitation of our prescription. 
We impose a Fourier-space turbulence spectrum rather than solving the MHD equations, and so our anisotropy is enforced relative to a single, fixed global field direction, rather than a local mean field that varies spatially throughout the volume, as in real MHD turbulence. 
As a result, our global treatment applies a similar perturbation across the entire field, producing comparable depolarization everywhere, rather than the local variation that arises in real MHD turbulence. 
This spatial coherence likely explains why our polarization fraction distributions are more tightly bound and less broad for sub-\alf cases than those from MHD simulations. Furthermore, for super-\alf cases, Cho-ENO is likely experiencing greater anisotropy on local scales, leading to smaller polarization fractions and a greater skew.

It is obvious that the models would differ in distribution based on their prescription, but what isn't is whether all three models would experience a positive shift in polarization skewness as a function of increasing \alf Mach number. Although the differences vary, for all three models, we see a shift from right to left hand distributions - a negative to positive skew trend. We can therefore confirm that the polarization skewness trend is not local to our model.

{
\small
\bibliographystyle{chicago}
\bibliography{./refs}

@ARTICLE{Ewertowski+2013,
       author = {{Ewertowski}, Bartek and {Basu}, Shantanu},
        title = "{A Mathematical Model for an Hourglass Magnetic Field}",
      journal = {\apj},
         year = 2013,
        month = apr,
       volume = {767},
       number = {1},
          eid = {33},
        pages = {33},
          doi = {10.1088/0004-637X/767/1/33},
archivePrefix = {arXiv},
       eprint = {1302.6913},
 primaryClass = {astro-ph.GA},
       adsurl = {https://ui.adsabs.harvard.edu/abs/2013ApJ...767...33E}
}

@ARTICLE{Cho+2003,
       author = {{Cho}, Jungyeon and {Lazarian}, A.},
        title = "{Compressible magnetohydrodynamic turbulence: mode coupling, scaling relations, anisotropy, viscosity-damped regime and astrophysical implications}",
      journal = {\mnras},
         year = 2003,
        month = oct,
       volume = {345},
       number = {12},
        pages = {325-339},
          doi = {10.1046/j.1365-8711.2003.06941.x},
archivePrefix = {arXiv},
       eprint = {astro-ph/0301062},
 primaryClass = {astro-ph},
       adsurl = {https://ui.adsabs.harvard.edu/abs/2003MNRAS.345..325C}
}

@ARTICLE{Lazarian+2006,
       author = {{Lazarian}, A.},
        title = "{Enhancement and Suppression of Heat Transfer by MHD Turbulence}",
      journal = {\apjl},
         year = 2006,
        month = jul,
       volume = {645},
       number = {1},
        pages = {L25-L28},
          doi = {10.1086/505796},
archivePrefix = {arXiv},
       eprint = {astro-ph/0608045},
 primaryClass = {astro-ph},
       adsurl = {https://ui.adsabs.harvard.edu/abs/2006ApJ...645L..25L}
}

@ARTICLE{Goldreich+1995,
       author = {{Goldreich}, P. and {Sridhar}, S.},
        title = "{Toward a Theory of Interstellar Turbulence. II. Strong Alfvenic Turbulence}",
      journal = {\apj},
         year = 1995,
        month = jan,
       volume = {438},
        pages = {763},
          doi = {10.1086/175121},
       adsurl = {https://ui.adsabs.harvard.edu/abs/1995ApJ...438..763G}
}

@INPROCEEDINGS{Zweibel+1996,
       author = {{Zweibel}, E.~G.},
        title = "{Polarimetry and the Theory of the Galactic Magnetic Field}",
    booktitle = {Polarimetry of the Interstellar Medium},
         year = 1996,
       editor = {{Roberge}, Wayne G. and {Whittet}, Doug C.~B.},
       series = {Astronomical Society of the Pacific Conference Series},
       volume = {97},
        month = jan,
        pages = {486},
       adsurl = {https://ui.adsabs.harvard.edu/abs/1996ASPC...97..486Z}
}

@ARTICLE{KennicuttEvans+2012,
       author = {{Kennicutt}, Robert C. and {Evans}, Neal J.},
        title = "{Star Formation in the Milky Way and Nearby Galaxies}",
      journal = {\araa},
         year = 2012,
        month = sep,
       volume = {50},
        pages = {531-608},
          doi = {10.1146/annurev-astro-081811-125610},
archivePrefix = {arXiv},
       eprint = {1204.3552},
 primaryClass = {astro-ph.GA},
       adsurl = {https://ui.adsabs.harvard.edu/abs/2012ARA&A..50..531K}
}

@ARTICLE{Matthews+2001,
       author = {{Matthews}, Brenda C. and {Wilson}, Christine D. and {Fiege}, Jason D.},
        title = "{Magnetic Fields in Star-forming Molecular Clouds. II. The Depolarization Effect in the OMC-3 Filament of Orion A}",
      journal = {\apj},
         year = 2001,
        month = nov,
       volume = {562},
       number = {1},
        pages = {400-423},
          doi = {10.1086/323375},
archivePrefix = {arXiv},
       eprint = {astro-ph/0106394},
 primaryClass = {astro-ph},
       adsurl = {https://ui.adsabs.harvard.edu/abs/2001ApJ...562..400M}
}

@ARTICLE{Planck+2015,
       author = {{Planck Collaboration {et al. }}},
        title = "{Planck intermediate results. XIX. An overview of the polarized thermal emission from Galactic dust}",
      journal = {\aap},
         year = 2015,
        month = apr,
       volume = {576},
          eid = {A104},
        pages = {A104},
          doi = {10.1051/0004-6361/201424082},
archivePrefix = {arXiv},
       eprint = {1405.0871},
 primaryClass = {astro-ph.GA},
       adsurl = {https://ui.adsabs.harvard.edu/abs/2015A&A...576A.104P}
}

@ARTICLE{Stephens+2013,
       author = {{Stephens}, Ian W. and {Looney}, Leslie W. and {Kwon}, Woojin and {Hull}, Charles L.~H. and {Plambeck}, Richard L. and {Crutcher}, Richard M. and {Chapman}, Nicholas and {Novak}, Giles and {Davidson}, Jacqueline and {Vaillancourt}, John E. and {Shinnaga}, Hiroko and {Matthews}, Tristan},
        title = "{The Magnetic Field Morphology of the Class 0 Protostar L1157-mm}",
      journal = {\apjl},
         year = 2013,
        month = may,
       volume = {769},
       number = {1},
          eid = {L15},
        pages = {L15},
          doi = {10.1088/2041-8205/769/1/L15},
archivePrefix = {arXiv},
       eprint = {1304.6739},
 primaryClass = {astro-ph.SR},
       adsurl = {https://ui.adsabs.harvard.edu/abs/2013ApJ...769L..15S}
}

@ARTICLE{Wang+2020,
       author = {{Wang}, Jia-Wei and {Koch}, Patrick M. and {Galv{\'a}n-Madrid}, Roberto and {Lai}, Shih-Ping and {Liu}, Hauyu Baobab and {Lin}, Sheng-Jun and {Pattle}, Kate},
        title = "{Formation of the Hub-Filament System G33.92+0.11: Local Interplay between Gravity, Velocity, and Magnetic Field}",
      journal = {\apj},
         year = 2020,
        month = dec,
       volume = {905},
       number = {2},
          eid = {158},
        pages = {158},
          doi = {10.3847/1538-4357/abc74e},
archivePrefix = {arXiv},
       eprint = {2011.01555},
 primaryClass = {astro-ph.GA},
       adsurl = {https://ui.adsabs.harvard.edu/abs/2020ApJ...905..158W}
}

@ARTICLE{Seifried+2020,
       author = {{Seifried}, D. and {Walch}, S. and {Weis}, M. and {Reissl}, S. and {Soler}, J.~D. and {Klessen}, R.~S. and {Joshi}, P.~R.},
        title = "{From parallel to perpendicular - On the orientation of magnetic fields in molecular clouds}",
      journal = {\mnras},
         year = 2020,
        month = oct,
       volume = {497},
       number = {4},
        pages = {4196-4212},
          doi = {10.1093/mnras/staa2231},
archivePrefix = {arXiv},
       eprint = {2003.00017},
 primaryClass = {astro-ph.GA},
       adsurl = {https://ui.adsabs.harvard.edu/abs/2020MNRAS.497.4196S}
}

@ARTICLE{InoueInutsuka+2009,
       author = {{Inoue}, Tsuyoshi and {Inutsuka}, Shu-ichiro},
        title = "{Two-Fluid Magnetohydrodynamics Simulations of Converging H I Flows in the Interstellar Medium. II. Are Molecular Clouds Generated Directly from a Warm Neutral Medium?}",
      journal = {\apj},
         year = 2009,
        month = oct,
       volume = {704},
       number = {1},
        pages = {161-169},
          doi = {10.1088/0004-637X/704/1/161},
archivePrefix = {arXiv},
       eprint = {0908.3701},
 primaryClass = {astro-ph.GA},
       adsurl = {https://ui.adsabs.harvard.edu/abs/2009ApJ...704..161I}
}

@ARTICLE{ChandrasekharFermi+1953,
       author = {{Chandrasekhar}, S. and {Fermi}, E.},
        title = "{Magnetic Fields in Spiral Arms.}",
      journal = {\apj},
         year = 1953,
        month = jul,
       volume = {118},
        pages = {113},
          doi = {10.1086/145731},
       adsurl = {https://ui.adsabs.harvard.edu/abs/1953ApJ...118..113C}
}

@ARTICLE{Davis+1951,
       author = {{Davis}, Leverett},
        title = "{The Strength of Interstellar Magnetic Fields}",
      journal = {Physical Review},
         year = 1951,
        month = mar,
       volume = {81},
       number = {5},
        pages = {890-891},
          doi = {10.1103/PhysRev.81.890.2},
       adsurl = {https://ui.adsabs.harvard.edu/abs/1951PhRv...81..890D}
}

@ARTICLE{Ostriker+2001,
       author = {{Ostriker}, Eve C. and {Stone}, James M. and {Gammie}, Charles F.},
        title = "{Density, Velocity, and Magnetic Field Structure in Turbulent Molecular Cloud Models}",
      journal = {\apj},
         year = 2001,
        month = jan,
       volume = {546},
       number = {2},
        pages = {980-1005},
          doi = {10.1086/318290},
archivePrefix = {arXiv},
       eprint = {astro-ph/0008454},
 primaryClass = {astro-ph},
       adsurl = {https://ui.adsabs.harvard.edu/abs/2001ApJ...546..980O}
}

@ARTICLE{Serkowski+1958,
       author = {{Serkowski}, K.},
        title = "{Statistical Analysis of the Polarization and Reddening of the Double Cluster in Perseus}",
      journal = {\actaa},
         year = 1958,
        month = jan,
       volume = {8},
        pages = {135},
       adsurl = {https://ui.adsabs.harvard.edu/abs/1958AcA.....8..135S}
}

@ARTICLE{Mouschovias1976a,
       author = {{Mouschovias}, T. Ch.},
        title = "{Nonhomologous contraction and equilibria of self-gravitating, magnetic interstellar clouds embedded in an intercloud medium: star formation. I. Formulation of the problem and method of solution.}",
      journal = {\apj},
         year = 1976,
        month = jun,
       volume = {206},
        pages = {753-767},
          doi = {10.1086/154436},
       adsurl = {https://ui.adsabs.harvard.edu/abs/1976ApJ...206..753M}
}

@ARTICLE{Mouschovias1976b,
       author = {{Mouschovias}, T.~C.},
        title = "{Nonhomologous contraction and equilibria of self-gravitating, magnetic interstellar clouds embedded in an intercloud medium: star formation. II. Results.}",
      journal = {\apj},
         year = 1976,
        month = jul,
       volume = {207},
        pages = {141-158},
          doi = {10.1086/154478},
       adsurl = {https://ui.adsabs.harvard.edu/abs/1976ApJ...207..141M}
}

@ARTICLE{GalliShu1993a,
       author = {{Galli}, Daniele and {Shu}, Frank H.},
        title = "{Collapse of Magnetized Molecular Cloud Cores. I. Semianalytical Solution}",
      journal = {\apj},
         year = 1993,
        month = nov,
       volume = {417},
        pages = {220},
          doi = {10.1086/173305},
       adsurl = {https://ui.adsabs.harvard.edu/abs/1993ApJ...417..220G}
}

@ARTICLE{Goodman+2009,
       author = {{Goodman}, Alyssa A. and {Pineda}, Jaime E. and {Schnee}, Scott L.},
        title = "{The ``True'' Column Density Distribution in Star-Forming Molecular Clouds}",
      journal = {\apj},
         year = 2009,
        month = feb,
       volume = {692},
       number = {1},
        pages = {91-103},
          doi = {10.1088/0004-637X/692/1/91},
archivePrefix = {arXiv},
       eprint = {0806.3441},
 primaryClass = {astro-ph},
       adsurl = {https://ui.adsabs.harvard.edu/abs/2009ApJ...692...91G}
}

@ARTICLE{GalliShu1993b,
       author = {{Galli}, Daniele and {Shu}, Frank H.},
        title = "{Collapse of Magnetized Molecular Cloud Cores. II. Numerical Results}",
      journal = {\apj},
         year = 1993,
        month = nov,
       volume = {417},
        pages = {243},
          doi = {10.1086/173306},
       adsurl = {https://ui.adsabs.harvard.edu/abs/1993ApJ...417..243G}
}

@ARTICLE{Hoang+2024,
       author = {{Hoang}, Thiem and {Truong}, Bao},
        title = "{Probing 3D Magnetic Fields Using Thermal Dust Polarization and Grain Alignment Theory}",
      journal = {\apj},
         year = 2024,
        month = apr,
       volume = {965},
       number = {2},
          eid = {183},
        pages = {183},
          doi = {10.3847/1538-4357/ad2a56},
archivePrefix = {arXiv},
       eprint = {2310.17048},
 primaryClass = {astro-ph.GA},
       adsurl = {https://ui.adsabs.harvard.edu/abs/2024ApJ...965..183H}
}

@ARTICLE{Wardle+1974,
       author = {{Wardle}, J.~F.~C. and {Kronberg}, P.~P.},
        title = "{The linear polarization of quasi-stellar radio sources at 3.71 and 11.1 centimeters.}",
      journal = {\apj},
         year = 1974,
        month = jan,
       volume = {194},
          eid = {249},
        pages = {249},
          doi = {10.1086/153240},
       adsurl = {https://ui.adsabs.harvard.edu/abs/1974ApJ...194..249W}
}

@ARTICLE{Hu+2023,
       author = {{Hu}, Yue and {Lazarian}, A.},
        title = "{Probing 3D magnetic fields - I. Polarized dust emission}",
      journal = {\mnras},
         year = 2023,
        month = mar,
       volume = {519},
       number = {3},
        pages = {3736-3748},
          doi = {10.1093/mnras/stac3744},
archivePrefix = {arXiv},
       eprint = {2203.09745},
 primaryClass = {astro-ph.GA},
       adsurl = {https://ui.adsabs.harvard.edu/abs/2023MNRAS.519.3736H}
}

@ARTICLE{Panopoulou+2016,
       author = {{Panopoulou}, G.~V. and {Psaradaki}, I. and {Tassis}, K.},
        title = "{The magnetic field and dust filaments in the Polaris Flare}",
      journal = {\mnras},
         year = 2016,
        month = oct,
       volume = {462},
       number = {2},
        pages = {1517-1529},
          doi = {10.1093/mnras/stw1678},
archivePrefix = {arXiv},
       eprint = {1607.00005},
 primaryClass = {astro-ph.GA},
       adsurl = {https://ui.adsabs.harvard.edu/abs/2016MNRAS.462.1517P}
}

@ARTICLE{Goodman+1992,
       author = {{Goodman}, Alyssa A. and {Jones}, Terry J. and {Lada}, Elizabeth A. and {Myers}, Philip C.},
        title = "{The Structure of Magnetic Fields in Dark Clouds: Infrared Polarimetry in B216--217}",
      journal = {\apj},
         year = 1992,
        month = nov,
       volume = {399},
        pages = {108},
          doi = {10.1086/171907},
       adsurl = {https://ui.adsabs.harvard.edu/abs/1992ApJ...399..108G}
}

@ARTICLE{Burkhart+2012,
       author = {{Burkhart}, Blakesley and {Lazarian}, A. and {Gaensler}, B.~M.},
        title = "{Properties of Interstellar Turbulence from Gradients of Linear Polarization Maps}",
      journal = {\apj},
         year = 2012,
        month = apr,
       volume = {749},
       number = {2},
          eid = {145},
        pages = {145},
          doi = {10.1088/0004-637X/749/2/145},
archivePrefix = {arXiv},
       eprint = {1111.3544},
 primaryClass = {astro-ph.GA},
       adsurl = {https://ui.adsabs.harvard.edu/abs/2012ApJ...749..145B}
}

@ARTICLE{Falceta+2008,
       author = {{Falceta-Gon{\c{c}}alves}, Diego and {Lazarian}, Alex and {Kowal}, Grzegorz},
        title = "{Studies of Regular and Random Magnetic Fields in the ISM: Statistics of Polarization Vectors and the Chandrasekhar-Fermi Technique}",
      journal = {\apj},
         year = 2008,
        month = may,
       volume = {679},
       number = {1},
        pages = {537-551},
          doi = {10.1086/587479},
archivePrefix = {arXiv},
       eprint = {0801.0279},
 primaryClass = {astro-ph},
       adsurl = {https://ui.adsabs.harvard.edu/abs/2008ApJ...679..537F}
}

@ARTICLE{Pattle+2017,
       author = {{Pattle}, Kate and {Ward-Thompson}, Derek and {Berry}, David and {Hatchell}, Jennifer and {Chen}, Huei-Ru and {Pon}, Andy and {Koch}, Patrick M. and {Kwon}, Woojin and {Kim}, Jongsoo and {Bastien}, Pierre and {Cho}, Jungyeon and {Coud{\'e}}, Simon and {Di Francesco}, James and {Fuller}, Gary and {Furuya}, Ray S. and {Graves}, Sarah F. and {Johnstone}, Doug and {Kirk}, Jason and {Kwon}, Jungmi and {Lee}, Chang Won and {Matthews}, Brenda C. and {Mottram}, Joseph C. and {Parsons}, Harriet and {Sadavoy}, Sarah and {Shinnaga}, Hiroko and {Soam}, Archana and {Hasegawa}, Tetsuo and {Lai}, Shih-Ping and {Qiu}, Keping and {Friberg}, Per},
        title = "{The JCMT BISTRO Survey: The Magnetic Field Strength in the Orion A Filament}",
      journal = {\apj},
         year = 2017,
        month = sep,
       volume = {846},
       number = {2},
          eid = {122},
        pages = {122},
          doi = {10.3847/1538-4357/aa80e5},
archivePrefix = {arXiv},
       eprint = {1707.05269},
 primaryClass = {astro-ph.GA},
       adsurl = {https://ui.adsabs.harvard.edu/abs/2017ApJ...846..122P}
}

@ARTICLE{Tahani+2018,
       author = {{Tahani}, M. and {Plume}, R. and {Brown}, J.~C. and {Kainulainen}, J.},
        title = "{Helical magnetic fields in molecular clouds?. A new method to determine the line-of-sight magnetic field structure in molecular clouds}",
      journal = {\aap},
         year = 2018,
        month = jun,
       volume = {614},
          eid = {A100},
        pages = {A100},
          doi = {10.1051/0004-6361/201732219},
archivePrefix = {arXiv},
       eprint = {1802.07831},
 primaryClass = {astro-ph.GA},
       adsurl = {https://ui.adsabs.harvard.edu/abs/2018A&A...614A.100T}
}

@ARTICLE{Stutz+2016,
       author = {{Stutz}, Amelia M. and {Gould}, Andrew},
        title = "{Slingshot mechanism in Orion: Kinematic evidence for ejection of protostars by filaments}",
      journal = {\aap},
         year = 2016,
        month = may,
       volume = {590},
          eid = {A2},
        pages = {A2},
          doi = {10.1051/0004-6361/201527979},
archivePrefix = {arXiv},
       eprint = {1512.04944},
 primaryClass = {astro-ph.SR},
       adsurl = {https://ui.adsabs.harvard.edu/abs/2016A&A...590A...2S}
}

@ARTICLE{Schleicher+2018,
       author = {{Schleicher}, Dominik R.~G. and {Stutz}, Amelia},
        title = "{Magnetic tension and instabilities in the Orion A integral-shaped filament}",
      journal = {\mnras},
         year = 2018,
        month = mar,
       volume = {475},
       number = {1},
        pages = {121-127},
          doi = {10.1093/mnras/stx2975},
archivePrefix = {arXiv},
       eprint = {1705.06302},
 primaryClass = {astro-ph.GA},
       adsurl = {https://ui.adsabs.harvard.edu/abs/2018MNRAS.475..121S}
}

@ARTICLE{Boekholt+2017,
       author = {{Boekholt}, T.~C.~N. and {Stutz}, A.~M. and {Fellhauer}, M. and {Schleicher}, D.~R.~G. and {Matus Carrillo}, D.~R.},
        title = "{Dynamical ejections of stars due to an accelerating gas filament}",
      journal = {\mnras},
         year = 2017,
        month = nov,
       volume = {471},
       number = {3},
        pages = {3590-3598},
          doi = {10.1093/mnras/stx1821},
archivePrefix = {arXiv},
       eprint = {1704.00720},
 primaryClass = {astro-ph.GA},
       adsurl = {https://ui.adsabs.harvard.edu/abs/2017MNRAS.471.3590B}
}

@ARTICLE{Reissl+2018,
       author = {{Reissl}, Stefan and {Stutz}, Amelia M. and {Brauer}, Robert and {Pellegrini}, Eric W. and {Schleicher}, Dominik R.~G. and {Klessen}, Ralf S.},
        title = "{Magnetic fields in star-forming systems (I): idealized synthetic signatures of dust polarization and Zeeman splitting in filaments}",
      journal = {\mnras},
         year = 2018,
        month = dec,
       volume = {481},
       number = {2},
        pages = {2507-2522},
          doi = {10.1093/mnras/sty2415},
archivePrefix = {arXiv},
       eprint = {1805.02674},
 primaryClass = {astro-ph.GA},
       adsurl = {https://ui.adsabs.harvard.edu/abs/2018MNRAS.481.2507R}
}

@ARTICLE{Scalo+1986,
       author = {{Scalo}, J.~M.},
        title = "{The Stellar Initial Mass Function}",
      journal = {\fcp},
         year = 1986,
        month = may,
       volume = {11},
        pages = {1-278},
       adsurl = {https://ui.adsabs.harvard.edu/abs/1986FCPh...11....1S}
}

@ARTICLE{Ballesteros+2007,
       author = {{Ballesteros-Paredes}, J. and {Hartmann}, L.},
        title = "{Remarks on Rapid vs. Slow Star Formation}",
      journal = {\rmxaa},
         year = 2007,
        month = apr,
       volume = {43},
        pages = {123-136},
archivePrefix = {arXiv},
       eprint = {astro-ph/0605268},
 primaryClass = {astro-ph},
       adsurl = {https://ui.adsabs.harvard.edu/abs/2007RMxAA..43..123B}
}

@ARTICLE{Palmeirim+2013,
       author = {{Palmeirim}, P. and {Andr{\'e}}, Ph. and {Kirk}, J. and {Ward-Thompson}, D. and {Arzoumanian}, D. and {K{\"o}nyves}, V. and {Didelon}, P. and {Schneider}, N. and {Benedettini}, M. and {Bontemps}, S. and {Di Francesco}, J. and {Elia}, D. and {Griffin}, M. and {Hennemann}, M. and {Hill}, T. and {Martin}, P.~G. and {Men'shchikov}, A. and {Molinari}, S. and {Motte}, F. and {Nguyen Luong}, Q. and {Nutter}, D. and {Peretto}, N. and {Pezzuto}, S. and {Roy}, A. and {Rygl}, K.~L.~J. and {Spinoglio}, L. and {White}, G.~L.},
        title = "{Herschel view of the Taurus B211/3 filament and striations: evidence of filamentary growth?}",
      journal = {\aap},
         year = 2013,
        month = feb,
       volume = {550},
          eid = {A38},
        pages = {A38},
          doi = {10.1051/0004-6361/201220500},
archivePrefix = {arXiv},
       eprint = {1211.6360},
 primaryClass = {astro-ph.SR},
       adsurl = {https://ui.adsabs.harvard.edu/abs/2013A&A...550A..38P}
}

@ARTICLE{Crutcher+2012,
       author = {{Crutcher}, Richard M.},
        title = "{Magnetic Fields in Molecular Clouds}",
      journal = {\araa},
         year = 2012,
        month = sep,
       volume = {50},
        pages = {29-63},
          doi = {10.1146/annurev-astro-081811-125514},
       adsurl = {https://ui.adsabs.harvard.edu/abs/2012ARA&A..50...29C}
}

@ARTICLE{Crutcher+2009,
       author = {{Crutcher}, Richard M. and {Hakobian}, Nicholas and {Troland}, Thomas H.},
        title = "{Testing Magnetic Star Formation Theory}",
      journal = {\apj},
         year = 2009,
        month = feb,
       volume = {692},
       number = {1},
        pages = {844-855},
          doi = {10.1088/0004-637X/692/1/844},
archivePrefix = {arXiv},
       eprint = {0807.2862},
 primaryClass = {astro-ph},
       adsurl = {https://ui.adsabs.harvard.edu/abs/2009ApJ...692..844C}
}

@ARTICLE{Shu+1987,
       author = {{Shu}, Frank H. and {Adams}, Fred C. and {Lizano}, Susana},
        title = "{Star formation in molecular clouds: observation and theory.}",
      journal = {\araa},
         year = 1987,
        month = jan,
       volume = {25},
        pages = {23-81},
          doi = {10.1146/annurev.aa.25.090187.000323},
       adsurl = {https://ui.adsabs.harvard.edu/abs/1987ARA&A..25...23S}
}

@ARTICLE{Soler+2017a,
       author = {{Soler}, J.~D. and {Ade}, P.~A.~R. and {Angil{\`e}}, F.~E. and {Ashton}, P. and {Benton}, S.~J. and {Devlin}, M.~J. and {Dober}, B. and {Fissel}, L.~M. and {Fukui}, Y. and {Galitzki}, N. and {Gandilo}, N.~N. and {Hennebelle}, P. and {Klein}, J. and {Li}, Z.-Y. and {Korotkov}, A.~L. and {Martin}, P.~G. and {Matthews}, T.~G. and {Moncelsi}, L. and {Netterfield}, C.~B. and {Novak}, G. and {Pascale}, E. and {Poidevin}, F. and {Santos}, F.~P. and {Savini}, G. and {Scott}, D. and {Shariff}, J.~A. and {Thomas}, N.~E. and {Tucker}, C.~E. and {Tucker}, G.~S. and {Ward-Thompson}, D.},
        title = "{The relation between the column density structures and the magnetic field orientation in the Vela C molecular complex}",
      journal = {\aap},
         year = 2017,
        month = jul,
       volume = {603},
          eid = {A64},
        pages = {A64},
          doi = {10.1051/0004-6361/201730608},
archivePrefix = {arXiv},
       eprint = {1702.03853},
 primaryClass = {astro-ph.GA},
       adsurl = {https://ui.adsabs.harvard.edu/abs/2017A&A...603A..64S}
}

@ARTICLE{Soler+2017b,
       author = {{Soler}, J.~D. and {Hennebelle}, P.},
        title = "{What are we learning from the relative orientation between density structures and the magnetic field in molecular clouds?}",
      journal = {\aap},
         year = 2017,
        month = oct,
       volume = {607},
          eid = {A2},
        pages = {A2},
          doi = {10.1051/0004-6361/201731049},
archivePrefix = {arXiv},
       eprint = {1705.00477},
 primaryClass = {astro-ph.GA},
       adsurl = {https://ui.adsabs.harvard.edu/abs/2017A&A...607A...2S}
}

@ARTICLE{Ferriere+2001,
       author = {{Ferri{\`e}re}, Katia M.},
        title = "{The interstellar environment of our galaxy}",
      journal = {Reviews of Modern Physics},
         year = 2001,
        month = oct,
       volume = {73},
       number = {4},
        pages = {1031-1066},
          doi = {10.1103/RevModPhys.73.1031},
archivePrefix = {arXiv},
       eprint = {astro-ph/0106359},
 primaryClass = {astro-ph},
       adsurl = {https://ui.adsabs.harvard.edu/abs/2001RvMP...73.1031F}
}

@ARTICLE{Shane+2024,
       author = {{Shane}, Brandon and {Burkhart}, Blakesley and {Fissel}, Laura and {Clark}, Susan E. and {Mocz}, Philip and {Foley}, Michael M.},
        title = "{Tracing 3-D Magnetic Field Structure Using Dust Polarization and the Zeeman Effect}",
      journal = {arXiv e-prints},
         year = 2024,
        month = nov,
          eid = {arXiv:2411.10286},
        pages = {arXiv:2411.10286},
          doi = {10.48550/arXiv.2411.10286},
archivePrefix = {arXiv},
       eprint = {2411.10286},
 primaryClass = {astro-ph.GA},
       adsurl = {https://ui.adsabs.harvard.edu/abs/2024arXiv241110286S}
}

@ARTICLE{Basu+2024,
       author = {{Basu}, Shantanu and {Li}, Xiyuan and {Bino}, Gianfranco},
        title = "{Hourglass Magnetic Field of a Protostellar System}",
      journal = {Universe},
         year = 2024,
        month = may,
       volume = {10},
       number = {5},
          eid = {218},
        pages = {218},
          doi = {10.3390/universe10050218},
archivePrefix = {arXiv},
       eprint = {2405.11069},
 primaryClass = {astro-ph.SR},
       adsurl = {https://ui.adsabs.harvard.edu/abs/2024Univ...10..218B}
}

@ARTICLE{Crutcher+1999,
       author = {{Crutcher}, Richard M.},
        title = "{Magnetic Fields in Molecular Clouds: Observations Confront Theory}",
      journal = {\apj},
         year = 1999,
        month = aug,
       volume = {520},
       number = {2},
        pages = {706-713},
          doi = {10.1086/307483},
       adsurl = {https://ui.adsabs.harvard.edu/abs/1999ApJ...520..706C}
}

@ARTICLE{Chen+2019,
       author = {{Chen}, Che-Yu and {King}, Patrick K. and {Li}, Zhi-Yun and {Fissel}, Laura M. and {Mazzei}, Renato R.},
        title = "{A new method to trace three-dimensional magnetic field structure within molecular clouds using dust polarization}",
      journal = {\mnras},
         year = 2019,
        month = may,
       volume = {485},
       number = {3},
        pages = {3499-3513},
          doi = {10.1093/mnras/stz618},
archivePrefix = {arXiv},
       eprint = {1810.10020},
 primaryClass = {astro-ph.GA},
       adsurl = {https://ui.adsabs.harvard.edu/abs/2019MNRAS.485.3499C}
}

@ARTICLE{Myers+2020,
       author = {{Myers}, Philip C. and {Stephens}, Ian W. and {Auddy}, Sayantan and {Basu}, Shantanu and {Bourke}, Tyler L. and {Hull}, Charles L.~H.},
        title = "{Magnetic Field Structure in Spheroidal Star-forming Clouds. II. Estimating Field Structure from Observed Maps}",
      journal = {\apj},
         year = 2020,
        month = jun,
       volume = {896},
       number = {2},
          eid = {163},
        pages = {163},
          doi = {10.3847/1538-4357/ab9110},
archivePrefix = {arXiv},
       eprint = {2005.04307},
 primaryClass = {astro-ph.GA},
       adsurl = {https://ui.adsabs.harvard.edu/abs/2020ApJ...896..163M}
}

@ARTICLE{Bonazzola1987,
       author = {{Bonazzola}, S. and {Heyvaerts}, J. and {Falgarone}, E. and {Perault}, M. and {Puget}, J.~L.},
        title = "{Jeans collapse in a turbulent medium}",
      journal = {\aap},
         year = 1987,
        month = jan,
       volume = {172},
       number = {1-2},
        pages = {293-298},
       adsurl = {https://ui.adsabs.harvard.edu/abs/1987A&A...172..293B}
}

@ARTICLE{Hildebrand+1988,
       author = {{Hildebrand}, Roger H.},
        title = "{Magnetic fields and stardust}",
      journal = {\qjras},
         year = 1988,
        month = sep,
       volume = {29},
        pages = {327-351},
       adsurl = {https://ui.adsabs.harvard.edu/abs/1988QJRAS..29..327H}
}

@ARTICLE{Bialy+2020,
       author = {{Bialy}, Shmuel and {Burkhart}, Blakesley},
        title = "{The Driving Scale-Density Decorrelation Scale Relation in a Turbulent Medium}",
      journal = {\apjl},
         year = 2020,
        month = may,
       volume = {894},
       number = {1},
          eid = {L2},
        pages = {L2},
          doi = {10.3847/2041-8213/ab8a32},
archivePrefix = {arXiv},
       eprint = {2001.06023},
 primaryClass = {astro-ph.GA},
       adsurl = {https://ui.adsabs.harvard.edu/abs/2020ApJ...894L...2B}
}

@ARTICLE{Grobschedl+2018,
       author = {{Gro{\ss}schedl}, Josefa E. and {Alves}, Jo{\~a}o and {Meingast}, Stefan and {Ackerl}, Christine and {Ascenso}, Joana and {Bouy}, Herv{\'e} and {Burkert}, Andreas and {Forbrich}, Jan and {F{\"u}rnkranz}, Verena and {Goodman}, Alyssa and {Hacar}, {\'A}lvaro and {Herbst-Kiss}, Gabor and {Lada}, Charles J. and {Larreina}, Irati and {Leschinski}, Kieran and {Lombardi}, Marco and {Moitinho}, Andr{\'e} and {Mortimer}, Daniel and {Zari}, Eleonora},
        title = "{3D shape of Orion A from Gaia DR2}",
      journal = {\aap},
         year = 2018,
        month = nov,
       volume = {619},
          eid = {A106},
        pages = {A106},
          doi = {10.1051/0004-6361/201833901},
archivePrefix = {arXiv},
       eprint = {1808.05952},
 primaryClass = {astro-ph.GA},
       adsurl = {https://ui.adsabs.harvard.edu/abs/2018A&A...619A.106G}
}

@ARTICLE{Portillo+2018,
       author = {{Portillo}, Stephen K.~N. and {Slepian}, Zachary and {Burkhart}, Blakesley and {Kahraman}, Sule and {Finkbeiner}, Douglas P.},
        title = "{Developing the 3-point Correlation Function for the Turbulent Interstellar Medium}",
      journal = {\apj},
         year = 2018,
        month = aug,
       volume = {862},
       number = {2},
          eid = {119},
        pages = {119},
          doi = {10.3847/1538-4357/aacb80},
archivePrefix = {arXiv},
       eprint = {1711.09907},
 primaryClass = {astro-ph.CO},
       adsurl = {https://ui.adsabs.harvard.edu/abs/2018ApJ...862..119P}
}

@ARTICLE{Burkhart+2020,
       author = {{Burkhart}, B. and {Appel}, S.~M. and {Bialy}, S. and {Cho}, J. and {Christensen}, A.~J. and {Collins}, D. and {Federrath}, C. and {Fielding}, D.~B. and {Finkbeiner}, D. and {Hill}, A.~S. and {Ib{\'a}{\~n}ez-Mej{\'\i}a}, J.~C. and {Krumholz}, M.~R. and {Lazarian}, A. and {Li}, M. and {Mocz}, P. and {Mac Low}, M.-M. and {Naiman}, J. and {Portillo}, S.~K.~N. and {Shane}, B. and {Slepian}, Z. and {Yuan}, Y.},
        title = "{The Catalogue for Astrophysical Turbulence Simulations (CATS)}",
      journal = {\apj},
         year = 2020,
        month = dec,
       volume = {905},
       number = {1},
          eid = {14},
        pages = {14},
          doi = {10.3847/1538-4357/abc484},
archivePrefix = {arXiv},
       eprint = {2010.11227},
 primaryClass = {astro-ph.GA},
       adsurl = {https://ui.adsabs.harvard.edu/abs/2020ApJ...905...14B}
}

@ARTICLE{Burkhart+2009,
       author = {{Burkhart}, Blakesley and {Falceta-Gon{\c{c}}alves}, D. and {Kowal}, G. and {Lazarian}, A.},
        title = "{Density Studies of MHD Interstellar Turbulence: Statistical Moments, Correlations and Bispectrum}",
      journal = {\apj},
         year = 2009,
        month = mar,
       volume = {693},
       number = {1},
        pages = {250-266},
          doi = {10.1088/0004-637X/693/1/250},
archivePrefix = {arXiv},
       eprint = {0811.0822},
 primaryClass = {astro-ph},
       adsurl = {https://ui.adsabs.harvard.edu/abs/2009ApJ...693..250B}
}

@ARTICLE{Hiltner+1949,
       author = {{Hiltner}, W.~A.},
        title = "{Polarization of Light from Distant Stars by Interstellar Medium}",
      journal = {Science},
         year = 1949,
        month = feb,
       volume = {109},
       number = {2825},
        pages = {165},
          doi = {10.1126/science.109.2825.165},
       adsurl = {https://ui.adsabs.harvard.edu/abs/1949Sci...109..165H}
}

@ARTICLE{Hall+1949,
       author = {{Hall}, John S.},
        title = "{Observations of the Polarized Light from Stars}",
      journal = {Science},
         year = 1949,
        month = feb,
       volume = {109},
       number = {2825},
        pages = {166-167},
          doi = {10.1126/science.109.2825.166},
       adsurl = {https://ui.adsabs.harvard.edu/abs/1949Sci...109..166H}
}

@ARTICLE{Hennebelle+2013,
       author = {{Hennebelle}, Patrick},
        title = "{On the origin of non-self-gravitating filaments in the ISM}",
      journal = {\aap},
         year = 2013,
        month = aug,
       volume = {556},
          eid = {A153},
        pages = {A153},
          doi = {10.1051/0004-6361/201321292},
archivePrefix = {arXiv},
       eprint = {1306.5452},
 primaryClass = {astro-ph.GA},
       adsurl = {https://ui.adsabs.harvard.edu/abs/2013A&A...556A.153H}
}

@ARTICLE{Hennebelle2008,
       author = {{Hennebelle}, Patrick and {Chabrier}, Gilles},
        title = "{Analytical Theory for the Initial Mass Function: CO Clumps and Prestellar Cores}",
      journal = {\apj},
         year = 2008,
        month = sep,
       volume = {684},
       number = {1},
        pages = {395-410},
          doi = {10.1086/589916},
archivePrefix = {arXiv},
       eprint = {0805.0691},
 primaryClass = {astro-ph},
       adsurl = {https://ui.adsabs.harvard.edu/abs/2008ApJ...684..395H}
}

@ARTICLE{Krumholz2005,
       author = {{Krumholz}, Mark R. and {McKee}, Christopher F.},
        title = "{A General Theory of Turbulence-regulated Star Formation, from Spirals to Ultraluminous Infrared Galaxies}",
      journal = {\apj},
         year = 2005,
        month = sep,
       volume = {630},
       number = {1},
        pages = {250-268},
          doi = {10.1086/431734},
archivePrefix = {arXiv},
       eprint = {astro-ph/0505177},
 primaryClass = {astro-ph},
       adsurl = {https://ui.adsabs.harvard.edu/abs/2005ApJ...630..250K}
}

@ARTICLE{Bonazzola1992,
       author = {{Bonazzola}, S. and {Perault}, M. and {Puget}, J.~L. and {Heyvaerts}, J. and {Falgarone}, E. and {Panis}, J.~F.},
        title = "{Jeans collapse of turbulent gas clouds - Tentative theory}",
      journal = {Journal of Fluid Mechanics},
         year = 1992,
        month = dec,
       volume = {245},
        pages = {1-28},
          doi = {10.1017/S0022112092000326},
       adsurl = {https://ui.adsabs.harvard.edu/abs/1992JFM...245....1B}
}

@ARTICLE{McKee2007,
       author = {{McKee}, Christopher F. and {Ostriker}, Eve C.},
        title = "{Theory of Star Formation}",
      journal = {\araa},
         year = 2007,
        month = sep,
       volume = {45},
       number = {1},
        pages = {565-687},
          doi = {10.1146/annurev.astro.45.051806.110602},
archivePrefix = {arXiv},
       eprint = {0707.3514},
 primaryClass = {astro-ph},
       adsurl = {https://ui.adsabs.harvard.edu/abs/2007ARA&A..45..565M}
}

@INPROCEEDINGS{Ballesteros2007,
       author = {{Ballesteros-Paredes}, J. and {Klessen}, R.~S. and {Mac Low}, M. -M. and {Vazquez-Semadeni}, E.},
        title = "{Molecular Cloud Turbulence and Star Formation}",
    booktitle = {Protostars and Planets V},
         year = 2007,
       editor = {{Reipurth}, Bo and {Jewitt}, David and {Keil}, Klaus},
        month = jan,
        pages = {63},
          doi = {10.48550/arXiv.astro-ph/0603357},
archivePrefix = {arXiv},
       eprint = {astro-ph/0603357},
 primaryClass = {astro-ph},
       adsurl = {https://ui.adsabs.harvard.edu/abs/2007prpl.conf...63B}
}

@ARTICLE{Maclow2004,
       author = {{Mac Low}, Mordecai-Mark and {Klessen}, Ralf S.},
        title = "{Control of star formation by supersonic turbulence}",
      journal = {Reviews of Modern Physics},
         year = 2004,
        month = jan,
       volume = {76},
       number = {1},
        pages = {125-194},
          doi = {10.1103/RevModPhys.76.125},
archivePrefix = {arXiv},
       eprint = {astro-ph/0301093},
 primaryClass = {astro-ph},
       adsurl = {https://ui.adsabs.harvard.edu/abs/2004RvMP...76..125M}
}

@ARTICLE{Adams87,
       author = {{Adams}, Fred C. and {Lada}, Charles J. and {Shu}, Frank H.},
        title = "{Spectral Evolution of Young Stellar Objects}",
      journal = {\apj},
         year = 1987,
        month = jan,
       volume = {312},
        pages = {788},
          doi = {10.1086/164924},
       adsurl = {https://ui.adsabs.harvard.edu/abs/1987ApJ...312..788A}
}

@ARTICLE{Ward+2017,
       author = {{Ward-Thompson}, Derek and {Pattle}, Kate and {Bastien}, Pierre and {Furuya}, Ray S. and {Kwon}, Woojin and {Lai}, Shih-Ping and {Qiu}, Keping and {Berry}, David and {Choi}, Minho and {Coud{\'e}}, Simon and {Di Francesco}, James and {Hoang}, Thiem and {Franzmann}, Erica and {Friberg}, Per and {Graves}, Sarah F. and {Greaves}, Jane S. and {Houde}, Martin and {Johnstone}, Doug and {Kirk}, Jason M. and {Koch}, Patrick M. and {Kwon}, Jungmi and {Lee}, Chang Won and {Li}, Di and {Matthews}, Brenda C. and {Mottram}, Joseph C. and {Parsons}, Harriet and {Pon}, Andy and {Rao}, Ramprasad and {Rawlings}, Mark and {Shinnaga}, Hiroko and {Sadavoy}, Sarah and {van Loo}, Sven and {Aso}, Yusuke and {Byun}, Do-Young and {Eswaraiah}, Chakali and {Chen}, Huei-Ru and {Chen}, Mike C.-Y. and {Chen}, Wen Ping and {Ching}, Tao-Chung and {Cho}, Jungyeon and {Chrysostomou}, Antonio and {Chung}, Eun Jung and {Doi}, Yasuo and {Drabek-Maunder}, Emily and {Eyres}, Stewart P.~S. and {Fiege}, Jason and {Friesen}, Rachel K. and {Fuller}, Gary and {Gledhill}, Tim and {Griffin}, Matt J. and {Gu}, Qilao and {Hasegawa}, Tetsuo and {Hatchell}, Jennifer and {Hayashi}, Saeko S. and {Holland}, Wayne and {Inoue}, Tsuyoshi and {Inutsuka}, Shu-ichiro and {Iwasaki}, Kazunari and {Jeong}, Il-Gyo and {Kang}, Ji-hyun and {Kang}, Miju and {Kang}, Sung-ju and {Kawabata}, Koji S. and {Kemper}, Francisca and {Kim}, Gwanjeong and {Kim}, Jongsoo and {Kim}, Kee-Tae and {Kim}, Kyoung Hee and {Kim}, Mi-Ryang and {Kim}, Shinyoung and {Lacaille}, Kevin M. and {Lee}, Jeong-Eun and {Lee}, Sang-Sung and {Li}, Dalei and {Li}, Hua-bai and {Liu}, Hong-Li and {Liu}, Junhao and {Liu}, Sheng-Yuan and {Liu}, Tie and {Lyo}, A.-Ran and {Mairs}, Steve and {Matsumura}, Masafumi and {Moriarty-Schieven}, Gerald H. and {Nakamura}, Fumitaka and {Nakanishi}, Hiroyuki and {Ohashi}, Nagayoshi and {Onaka}, Takashi and {Peretto}, Nicolas and {Pyo}, Tae-Soo and {Qian}, Lei and {Retter}, Brendan and {Richer}, John and {Rigby}, Andrew and {Robitaille}, Jean-Fran{\c{c}}ois and {Savini}, Giorgio and {Scaife}, Anna M.~M. and {Soam}, Archana and {Tamura}, Motohide and {Tang}, Ya-Wen and {Tomisaka}, Kohji and {Wang}, Hongchi and {Wang}, Jia-Wei and {Whitworth}, Anthony P. and {Yen}, Hsi-Wei and {Yoo}, Hyunju and {Yuan}, Jinghua and {Zhang}, Chuan-Peng and {Zhang}, Guoyin and {Zhou}, Jianjun and {Zhu}, Lei and {Andr{\'e}}, Philippe and {Dowell}, C. Darren and {Falle}, Sam and {Tsukamoto}, Yusuke},
        title = "{First Results from BISTRO: A SCUBA-2 Polarimeter Survey of the Gould Belt}",
      journal = {\apj},
         year = 2017,
        month = jun,
       volume = {842},
       number = {1},
          eid = {66},
        pages = {66},
          doi = {10.3847/1538-4357/aa70a0},
archivePrefix = {arXiv},
       eprint = {1704.08552},
 primaryClass = {astro-ph.GA},
       adsurl = {https://ui.adsabs.harvard.edu/abs/2017ApJ...842...66W}
}

@INPROCEEDINGS{Lizano87,
       author = {{Lizano}, Susana and {Shu}, Frank H.},
        title = "{Formation and heating of molecular cloud cores.}",
    booktitle = {Physical Processes in Interstellar Clouds},
         year = 1987,
       editor = {{Morfill}, G.~E. and {Scholer}, M.},
       series = {NATO Advanced Study Institute (ASI) Series C},
       volume = {210},
        month = jan,
        pages = {173-193},
          doi = {10.1007/978-94-009-3945-5_14},
       adsurl = {https://ui.adsabs.harvard.edu/abs/1987ASIC..210..173L}
}

@ARTICLE{Elmegreen2004,
       author = {{Elmegreen}, Bruce G. and {Scalo}, John},
        title = "{Interstellar Turbulence I: Observations and Processes}",
      journal = {\araa},
         year = 2004,
        month = sep,
       volume = {42},
       number = {1},
        pages = {211-273},
          doi = {10.1146/annurev.astro.41.011802.094859},
archivePrefix = {arXiv},
       eprint = {astro-ph/0404451},
 primaryClass = {astro-ph},
       adsurl = {https://ui.adsabs.harvard.edu/abs/2004ARA&A..42..211E}
}

@ARTICLE{Chuss2019,
       author = {{Chuss}, David T. and {Andersson}, B. -G. and {Bally}, John and {Dotson}, Jessie L. and {Dowell}, C. Darren and {Guerra}, Jordan A. and {Harper}, Doyal A. and {Houde}, Martin and {Jones}, Terry Jay and {Lazarian}, A. and {Lopez Rodriguez}, Enrique and {Michail}, Joseph M. and {Morris}, Mark R. and {Novak}, Giles and {Siah}, Javad and {Staguhn}, Johannes and {Vaillancourt}, John E. and {Volpert}, C.~G. and {Werner}, Michael and {Wollack}, Edward J. and {Benford}, Dominic J. and {Berthoud}, Marc and {Cox}, Erin G. and {Crutcher}, Richard and {Dale}, Daniel A. and {Fissel}, L.~M. and {Goldsmith}, Paul F. and {Hamilton}, Ryan T. and {Hanany}, Shaul and {Henning}, Thomas K. and {Looney}, Leslie W. and {Moseley}, S. Harvey and {Santos}, Fabio P. and {Stephens}, Ian and {Tassis}, Konstantinos and {Trinh}, Christopher Q. and {Van Camp}, Eric and {Ward-Thompson}, Derek and {HAWC + Science Team}},
        title = "{HAWC+/SOFIA Multiwavelength Polarimetric Observations of OMC-1}",
      journal = {\apj},
         year = 2019,
        month = feb,
       volume = {872},
       number = {2},
          eid = {187},
        pages = {187},
          doi = {10.3847/1538-4357/aafd37},
archivePrefix = {arXiv},
       eprint = {1810.08233},
 primaryClass = {astro-ph.GA},
       adsurl = {https://ui.adsabs.harvard.edu/abs/2019ApJ...872..187C}
}

@ARTICLE{SOFIA2018,
       author = {{Harper}, Doyal A. and {Runyan}, Marcus C. and {Dowell}, C. Darren and {Wirth}, C. Jesse and {Amato}, Michael and {Ames}, Troy and {Amiri}, Mandana and {Banks}, Stuart and {Bartels}, Arlin and {Benford}, Dominic J. and {Berthoud}, Marc and {Buchanan}, Ernest and {Casey}, Sean and {Chapman}, Nicholas L. and {Chuss}, David T. and {Cook}, Brant and {Derro}, Rebecca and {Dotson}, Jessie L. and {Evans}, Rhodri and {Fixsen}, Dale and {Gatley}, Ian and {Guerra}, Jordan A. and {Halpern}, Mark and {Hamilton}, Ryan T. and {Hamlin}, Louise A. and {Hansen}, Christopher J. and {Heimsath}, Stephen and {Hermida}, Alfonso and {Hilton}, Gene C. and {Hirsch}, Robert and {Hollister}, Matthew I. and {Hostetter}, Carl F. and {Irwin}, Kent and {Jhabvala}, Christine A. and {Jhabvala}, Murzban and {Kastner}, Joel and {Kov{\'a}cs}, Attila and {Lin}, Sean and {Loewenstein}, Robert F. and {Looney}, Leslie W. and {Lopez-Rodriguez}, Enrique and {Maher}, Stephen F. and {Michail}, Joseph M. and {Miller}, Timothy M. and {Moseley}, S. Harvey and {Novak}, Giles and {Pernic}, Robert J. and {Rennick}, Timothy and {Rhody}, Harvey and {Sandberg}, Eric and {Sandford}, Dale and {Santos}, Fabio Pereira and {Shafer}, Rick and {Sharp}, Elmer H. and {Shirron}, Peter and {Siah}, Javad and {Silverberg}, Robert and {Sparr}, Leroy M. and {Spotz}, Robert and {Staguhn}, Johannes G. and {Toorian}, Armen S. and {Towey}, Shannon and {Tuttle}, Jim and {Vaillancourt}, John and {Voellmer}, George and {Volpert}, Carolyn G. and {Wang}, Shu-I. and {Wollack}, Edward J.},
        title = "{HAWC+, the Far-Infrared Camera and Polarimeter for SOFIA}",
      journal = {Journal of Astronomical Instrumentation},
         year = 2018,
        month = jan,
       volume = {7},
       number = {4},
          eid = {1840008-1025},
        pages = {1840008-1025},
          doi = {10.1142/S2251171718400081},
       adsurl = {https://ui.adsabs.harvard.edu/abs/2018JAI.....740008H}
}

@ARTICLE{Hartmann2002,
       author = {{Hartmann}, Lee},
        title = "{Flows, Fragmentation, and Star Formation. I. Low-Mass Stars in Taurus}",
      journal = {\apj},
         year = 2002,
        month = oct,
       volume = {578},
       number = {2},
        pages = {914-924},
          doi = {10.1086/342657},
archivePrefix = {arXiv},
       eprint = {astro-ph/0207216},
 primaryClass = {astro-ph},
       adsurl = {https://ui.adsabs.harvard.edu/abs/2002ApJ...578..914H}
}

@ARTICLE{FiegePudritz2000a,
       author = {{Fiege}, Jason D. and {Pudritz}, Ralph E.},
        title = "{Helical fields and filamentary molecular clouds - I}",
      journal = {\mnras},
         year = 2000,
        month = jan,
       volume = {311},
       number = {1},
        pages = {85-104},
          doi = {10.1046/j.1365-8711.2000.03066.x},
archivePrefix = {arXiv},
       eprint = {astro-ph/9901096},
 primaryClass = {astro-ph},
       adsurl = {https://ui.adsabs.harvard.edu/abs/2000MNRAS.311...85F}
}

@ARTICLE{FiegePudritz2000b,
       author = {{Fiege}, Jason D. and {Pudritz}, Ralph E.},
        title = "{Helical fields and filamentary molecular clouds - II. Axisymmetric stability and fragmentation}",
      journal = {\mnras},
         year = 2000,
        month = jan,
       volume = {311},
       number = {1},
        pages = {105-119},
          doi = {10.1046/j.1365-8711.2000.03067.x},
archivePrefix = {arXiv},
       eprint = {astro-ph/9902385},
 primaryClass = {astro-ph},
       adsurl = {https://ui.adsabs.harvard.edu/abs/2000MNRAS.311..105F}
}

@ARTICLE{FiegePudritz2000c,
       author = {{Fiege}, Jason D. and {Pudritz}, Ralph E.},
        title = "{Polarized Submillimeter Emission from Filamentary Molecular Clouds}",
      journal = {\apj},
         year = 2000,
        month = dec,
       volume = {544},
       number = {2},
        pages = {830-837},
          doi = {10.1086/317228},
archivePrefix = {arXiv},
       eprint = {astro-ph/0005363},
 primaryClass = {astro-ph},
       adsurl = {https://ui.adsabs.harvard.edu/abs/2000ApJ...544..830F}
}

@ARTICLE{Heitsch+2009,
       author = {{Heitsch}, Fabian and {Stone}, James M. and {Hartmann}, Lee W.},
        title = "{Effects of Magnetic Field Strength and Orientation on Molecular Cloud Formation}",
      journal = {\apj},
         year = 2009,
        month = apr,
       volume = {695},
       number = {1},
        pages = {248-258},
          doi = {10.1088/0004-637X/695/1/248},
archivePrefix = {arXiv},
       eprint = {0812.3339},
 primaryClass = {astro-ph},
       adsurl = {https://ui.adsabs.harvard.edu/abs/2009ApJ...695..248H}
}

@ARTICLE{Heitsch+2001,
       author = {{Heitsch}, Fabian and {Zweibel}, Ellen G. and {Mac Low}, Mordecai-Mark and {Li}, Pakshing and {Norman}, Michael L.},
        title = "{Magnetic Field Diagnostics Based on Far-Infrared Polarimetry: Tests Using Numerical Simulations}",
      journal = {\apj},
         year = 2001,
        month = nov,
       volume = {561},
       number = {2},
        pages = {800-814},
          doi = {10.1086/323489},
archivePrefix = {arXiv},
       eprint = {astro-ph/0103286},
 primaryClass = {astro-ph},
       adsurl = {https://ui.adsabs.harvard.edu/abs/2001ApJ...561..800H}
}

@ARTICLE{Lazarian+1999,
       author = {{Lazarian}, A. and {Vishniac}, Ethan T.},
        title = "{Reconnection in a Weakly Stochastic Field}",
      journal = {\apj},
         year = 1999,
        month = jun,
       volume = {517},
       number = {2},
        pages = {700-718},
          doi = {10.1086/307233},
archivePrefix = {arXiv},
       eprint = {astro-ph/9811037},
 primaryClass = {astro-ph},
       adsurl = {https://ui.adsabs.harvard.edu/abs/1999ApJ...517..700L}
}

@ARTICLE{Qiu+2014,
       author = {{Qiu}, Keping and {Zhang}, Qizhou and {Menten}, Karl M. and {Liu}, Hauyu B. and {Tang}, Ya-Wen and {Girart}, Josep M.},
        title = "{Submillimeter Array Observations of Magnetic Fields in G240.31+0.07: An Hourglass in a Massive Cluster-forming Core}",
      journal = {\apjl},
         year = 2014,
        month = oct,
       volume = {794},
       number = {1},
          eid = {L18},
        pages = {L18},
          doi = {10.1088/2041-8205/794/1/L18},
archivePrefix = {arXiv},
       eprint = {1409.5608},
 primaryClass = {astro-ph.GA},
       adsurl = {https://ui.adsabs.harvard.edu/abs/2014ApJ...794L..18Q}
}

@ARTICLE{Girart+2006,
       author = {{Girart}, Josep M. and {Rao}, Ramprasad and {Marrone}, Daniel P.},
        title = "{Magnetic Fields in the Formation of Sun-Like Stars}",
      journal = {Science},
         year = 2006,
        month = aug,
       volume = {313},
       number = {5788},
        pages = {812-814},
          doi = {10.1126/science.1129093},
archivePrefix = {arXiv},
       eprint = {astro-ph/0609177},
 primaryClass = {astro-ph},
       adsurl = {https://ui.adsabs.harvard.edu/abs/2006Sci...313..812G}
}

@ARTICLE{Maron+2001,
       author = {{Maron}, Jason and {Goldreich}, Peter},
        title = "{Simulations of Incompressible Magnetohydrodynamic Turbulence}",
      journal = {\apj},
         year = 2001,
        month = jun,
       volume = {554},
       number = {2},
        pages = {1175-1196},
          doi = {10.1086/321413},
archivePrefix = {arXiv},
       eprint = {astro-ph/0012491},
 primaryClass = {astro-ph},
       adsurl = {https://ui.adsabs.harvard.edu/abs/2001ApJ...554.1175M}
}

@ARTICLE{Cho+2000,
       author = {{Cho}, Jungyeon and {Vishniac}, Ethan T.},
        title = "{The Anisotropy of Magnetohydrodynamic Alfv{\'e}nic Turbulence}",
      journal = {\apj},
         year = 2000,
        month = aug,
       volume = {539},
       number = {1},
        pages = {273-282},
          doi = {10.1086/309213},
archivePrefix = {arXiv},
       eprint = {astro-ph/0003403},
 primaryClass = {astro-ph},
       adsurl = {https://ui.adsabs.harvard.edu/abs/2000ApJ...539..273C}
}

@ARTICLE{Armstrong+1995,
       author = {{Armstrong}, J.~W. and {Rickett}, B.~J. and {Spangler}, S.~R.},
        title = "{Electron Density Power Spectrum in the Local Interstellar Medium}",
      journal = {\apj},
         year = 1995,
        month = apr,
       volume = {443},
        pages = {209},
          doi = {10.1086/175515},
       adsurl = {https://ui.adsabs.harvard.edu/abs/1995ApJ...443..209A}
}

@BOOK{Monin+1975,
       author = {{Monin}, A.~S. and {Iaglom}, A.~M.},
        title = "{Statistical fluid mechanics: Mechanics of turbulence. Volume 2 /revised and enlarged edition/}",
         year = 1975,
       adsurl = {https://ui.adsabs.harvard.edu/abs/1975mit..bookR....M}
}

@ARTICLE{Reissl+2021,
       author = {{Reissl}, Stefan and {Stutz}, Amelia M. and {Klessen}, Ralf S. and {Seifried}, Daniel and {Walch}, Stefanie},
        title = "{Magnetic fields in star-forming systems - II: Examining dust polarization, the Zeeman effect, and the Faraday rotation measure as magnetic field tracers}",
      journal = {\mnras},
         year = 2021,
        month = jan,
       volume = {500},
       number = {1},
        pages = {153-176},
          doi = {10.1093/mnras/staa3148},
archivePrefix = {arXiv},
       eprint = {2009.04201},
 primaryClass = {astro-ph.GA},
       adsurl = {https://ui.adsabs.harvard.edu/abs/2021MNRAS.500..153R}
}

@ARTICLE{Stone+2020,
       author = {{Stone}, James M. and {Tomida}, Kengo and {White}, Christopher J. and {Felker}, Kyle G.},
        title = "{The Athena++ Adaptive Mesh Refinement Framework: Design and Magnetohydrodynamic Solvers}",
      journal = {\apjs},
         year = 2020,
        month = jul,
       volume = {249},
       number = {1},
          eid = {4},
        pages = {4},
          doi = {10.3847/1538-4365/ab929b},
archivePrefix = {arXiv},
       eprint = {2005.06651},
 primaryClass = {astro-ph.IM},
       adsurl = {https://ui.adsabs.harvard.edu/abs/2020ApJS..249....4S}
}

@ARTICLE{MestelSpitzer1956,
       author = {{Mestel}, L. and {Spitzer}, Jr., L.},
        title = "{Star formation in magnetic dust clouds}",
      journal = {\mnras},
         year = 1956,
        month = jan,
       volume = {116},
        pages = {503},
          doi = {10.1093/mnras/116.5.503},
       adsurl = {https://ui.adsabs.harvard.edu/abs/1956MNRAS.116..503M}
}

@ARTICLE{Hartmannetal2001,
       author = {{Hartmann}, Lee and {Ballesteros-Paredes}, Javier and {Bergin}, Edwin A.},
        title = "{Rapid Formation of Molecular Clouds and Stars in the Solar Neighborhood}",
      journal = {\apj},
         year = 2001,
        month = dec,
       volume = {562},
       number = {2},
        pages = {852-868},
          doi = {10.1086/323863},
archivePrefix = {arXiv},
       eprint = {astro-ph/0108023},
 primaryClass = {astro-ph},
       adsurl = {https://ui.adsabs.harvard.edu/abs/2001ApJ...562..852H}
}
}

\end{document}